\magnification1200


\vskip 2cm
\centerline
{\bf  $E_{11}$  and exceptional field theory}
\vskip 1cm
\centerline{ Alexander G. Tumanov and Peter West}
\centerline{Department of Mathematics}
\centerline{King's College, London WC2R 2LS, UK}
\vskip 2cm
\leftline{\sl Abstract}
We demonstrate that exceptional field theory is a truncation of the non-linear realisation of the semi-direct product of $E_{11}$ and its first fundamental as proposed  in 2003.  Evaluating  the simple equations of the $E_{11}$ approach, and using  the commutators of the $E_{11}$ algebra,  we  find the equations of exceptional field theory after making a radical truncation. This proceedure does not respect any of  the higher level $E_{11}$ symmetries  and so these are lost. We suggest that the need for the section condition in exceptional field theory could be a consequence of the truncation. 
\vskip2cm
\noindent

\vskip .5cm

\vfill
\eject
{\bf 1 Introduction}
\medskip
The scalars that are part of a supergravity multiplet occur in the corresponding supergravity theory as a coset, or non-linear realisation,  and  the full supergravity theory possess the symmetries of the coset. This phenomenon was first observed in the context of the four dimensional $N=4$  supergravity [1]. However, it was a surprise when it was found that the maximal supergravity theory in four dimensions possessed  seventy scalars which belonged to a coset of $E_7$ with subgroup $SU(8)$ [2]. The other maximal supergravities in $D$ dimensions for $D\le 9$ were then found to possess a corresponding 
$E_{11-D}$ symmetry [3]. While the IIB supergravity theory posses a SL(2,R) symmetry [4]. 
\par
It was almost universally thought that the $E_{n}$ symmetries found in the maximal supergravity theories were a result of the dimensional reduction process and did not lift to the ten or eleven dimensional theories. As a result it was not thought that the symmetries were part of any  underlying theory. However in 2001 it was proposed that the underlying theory of strings and branes had an $E_{11}$ symmetry regardless of which 
dimension it was formulated in [5]. In particular, it was conjectured that the $E_{11}$ symmetry was encoded in a non-linear realisation which extended the known supergravity theories. This proposal was motivated by the fact [6] that the maximal supergravity theories could be formulated as  non-linear realisations of an algebra which contained traces of the $E_{11}$ algebra. This conjecture can be put another way; it was proposed that  the coset, or non-linear realisation, that the scalars belonged to generalised to include the other fields of the supergravity theory and that  the underlying group had the features  we are familiar in physics, that is, it is a Kac-Moody group, then   as a  consequence one  that one had include in the non-linear realisation many more fields in the theory than just those found in the usual supergravity theories.  This procedure  involved introducing generators that carried space-time indices, that is, Lorentz indices corresponding to the fields that were not scalars. 
An account of the literature on symmetries in supergravity theories  can be found in reference [5] and  an even more detailed account in [7]. 
\par
The paper of reference [5] put to one side the question of how to systematically incorporate  space-time into the theory, however, this question was addressed in 2003 when it was proposed that the one should consider the non-linear realisation of the semi-direct product of $E_{11}$ with generators that belonged to the first fundamental representation of $E_{11}$, called the $l_1$ representation [8].  While the $E_{11}$ part of the non-linear realisation lead to an infinite number of fields the first few of which were the fields of the maximal supergravity theories,  the $l_1$ part lead to an infinite number of coordinates that were in a one to one relation with the generators of the $l_1$ realisation. To be more precise, taking into account the local symmetries of the non-linear realisation, for each generator in the Borel subgroup of $E_{11}$ the theory possessed a corresponding field, while for each generator of the $l_1$ representation one found a coordinate. The fields then depended on the coordinates of this generalised space-time. 
\par 
The theories in the different dimensions emerge by carrying out the $E_{11}\otimes_s l_1 $ non-linear realisation for different decompositions of the algebra [9,10,11,12]. The theory in $D$ dimensions comes from taking the decomposition into $GL(D)\times E_{11-D}$. This choice is unique except for ten dimensions where  there are two possible choices of GL(10), or $A_9$  and so  two possible theories in ten dimensions [5,6]. These are, of course,  the IIA and IIB theories. In each dimension the generators of $E_{11}$  (fields) can be classified by a level which is the number of 
up (down) minus down (up) indices. The two exceptions are  in eleven dimensions where this number must be divided by three and the ten dimensional IIA theory where the counting  is straightforward,  but more subtile. The same is true for the generators (coordinates) of the $l_1$ representation but one must add (subtract) one.   
\par
There is very good evidence that the $l_1$ representation contains all  brane charges [8,13,14,15]. Consequently,    there is  a one to one correspondence between the coordinates of the generalised space-time and the brane charges. Indeed one can think of each coordinate as associated with a given type of brane probe.  Furthermore, for every generator  in the Borel
subalgebra  of $E_{11}$ there is at least one  element in the $l_1$
representation [13],  and  as a result   for every field at low level in
the non-linear realisation one finds at least one  corresponding coordinate. 
The reader will easily see this  correspondence between the coordinates and fields at low levels.  
Papers [15,14,13]  discussed the appearance in the $l_1$ representation of "exotic" representations of GL(D), that is, representations that are not just labelled by a single set of antisymmetrised indices. The appearance of  "exotic" brane charges implies the existence of correspondingly exotic generalised space-time coordinates. 
\par

In any dimension, with the exception of the IIA theory,  the  level zero coordinates of the $l_1$ representation are just  the  usual coordinates of space-time,  $x^a$. In eleven dimensions the $l_1$ representation leads to the coordinates [8,13]
$$
x^a (0), x_{ab}(1), x_{a_1\ldots a_5}(2), x_{a_1\ldots a_7,b}(3), x_{a_1\ldots a_8}(3),
$$
$$
\  x_{\hat b_1\hat b_2 \hat b_3, \hat a_1\ldots \hat a_8}\ (4),
\  x_{(\hat c \hat d ), \hat a_1\ldots \hat a_9}\ (4),
\  x_{\hat c\hat d,\hat a_1\ldots \hat a_9}\ (4),\ 
\  x_{\hat c,\hat a_1\ldots \hat a_{10}}\ (4),\ 
\ldots  
\eqno(1.1)$$
where the number in brackets gives the level. The coordinates up to and including level seven were given  in reference [13]. An analysis of the  equations determining the  generalised coordinates  in lower dimensions was given in reference [15].   At level one in  the $l_1 $ representation one finds coordinates which are 
 scalars under the Lorentz group but transform as   [15,16]
 
$$10,\quad\overline {16}, \quad \overline {27},\quad  56, \quad {\rm  and}\quad 
248\oplus 1 , \quad {\rm of }\quad SL(5),\quad  SO(5,5),\quad 
E_6, \quad  E_7\quad {\rm  and}\  E_8 
\eqno(1.2)$$
 for $  D= 7,6,5,4$ and $3$    dimensions  respectively.
\par
The generators of the $l_1$ representation in $D$ dimensions that are forms, that is, carry a single set of completely anti-symmetrised space-time indices,    are listed  [12,15,16,14] in the table  below. The coordinates of the generalised space-time, being in   one to one correspondence with the generators of the $l_1$ representation,  it is trivial to read them off from the table.  In
the first column we find the Lorentz scalar coordinates mentioned just above.  
\par 
A not well appreciated point is that the non-linear realisation does not just specify the representations of the fields and coordinates contained in  the theory, it also specifies the dynamics.  By its definition the non-linear realisation  possess specified symmetries which act on the   group element. The fields and coordinates of the theory are the  parameters in  the  group element and as the symmetries determine the way the group element appears in the theory they  also determine the way in which the fields appear.  
\par
Non-linear realisations played a crucial part in the development of particle physics.   The group of interest to researchers at that time was $SU(2)\times SU(2)$ with local subgroup being the diagonal subgroup SU(2).The  dynamics which the non-linear realisation determined was found to be that  of the pions and as a result researchers begun to understand the role symmetry played in the underlying theory.

\medskip
\bigskip
{\centerline{\bf {Table. The generators   in the $l_1$ representation  in D
dimensions}}}
\medskip
$$\halign{\centerline{#} \cr
\vbox{\offinterlineskip
\halign{\strut \vrule \quad \hfil # \hfil\quad &\vrule Ê\quad \hfil #
\hfil\quad &\vrule \hfil # \hfil
&\vrule \hfil # \hfil Ê&\vrule \hfil # \hfil &\vrule \hfil # \hfil &
\vrule \hfil # \hfil &\vrule \hfil # \hfil &\vrule \hfil # \hfil &
\vrule \hfil # \hfil &\vrule#
\cr
\noalign{\hrule}
D&G&$Z$&$Z^{a}$&$Z^{a_1a_2}$&$Z^{a_1\ldots a_{3}}$&$Z^{a_1\ldots a_
{4}}$&$Z^{a_1\ldots a_{5}}$&$Z^{a_1\ldots a_6}$&$Z^{a_1\ldots a_7}$&\cr
\noalign{\hrule}
8&$SL(3)\otimes SL(2)$&$\bf (3,2)$&$\bf (\bar 3,1)$&$\bf (1,2)$&$\bf
(3,1)$&$\bf (\bar 3,2)$&$\bf (1,3)$&$\bf (3,2)$&$\bf (6,1)$&\cr
&&&&&&&$\bf (8,1)$&$\bf (6,2)$&$\bf (18,1)$&\cr Ê&&&&&&&$\bf (1,1)$&&$
\bf
(3,1)$&\cr Ê&&&&&&&&&$\bf (6,1)$&\cr
&&&&&&&&&$\bf (3,3)$&\cr
\noalign{\hrule}
7&$SL(5)$&$\bf 10$&$\bf\bar 5$&$\bf 5$&$\bf \overline {10}$&$\bf 24$&$\bf
40$&$\bf 70$&-&\cr Ê&&&&&&$\bf 1$&$\bf 15$&$\bf 50$&-&\cr
&&&&&&&$\bf 10$&$\bf 45$&-&\cr
&&&&&&&&$\bf 5$&-&\cr
\noalign{\hrule}
6&$SO(5,5)$&$\bf \overline {16}$&$\bf 10$&$\bf 16$&$\bf 45$&$\bf \overline
{144}$&$\bf 320$&-&-&\cr &&&&&$\bf 1$&$\bf 16$&$\bf 126$&-&-&\cr
&&&&&&&$\bf 120$&-&-&\cr
\noalign{\hrule}
5&$E_6$&$\bf\overline { 27}$&$\bf 27$&$\bf 78$&$\bf \overline {351}$&$\bf
1728$&-&-&-&\cr Ê&&&&$\bf 1$&$\bf \overline {27}$&$\bf 351$&-&-&-&\cr
&&&&&&$\bf 27$&-&-&-&\cr
\noalign{\hrule}
4&$E_7$&$\bf 56$&$\bf 133$&$\bf 912$&$\bf 8645$&-&-&-&-&\cr
&&&$\bf 1$&$\bf 56$&$\bf 1539$&-&-&-&-&\cr
&&&&&$\bf 133$&-&-&-&-&\cr
&&&&&$\bf 1$&-&-&-&-&\cr
\noalign{\hrule}
3&$E_8$&$\bf 248$&$\bf 3875$&$\bf 147250$&-&-&-&-&-&\cr
&&$\bf1$&$\bf248$&$\bf 30380$&-&-&-&-&-&\cr
&&&$\bf 1$&$\bf 3875$&-&-&-&-&-&\cr
&&&&$\bf 248$&-&-&-&-&-&\cr
&&&&$\bf 1$&-&-&-&-&-&\cr
\noalign{\hrule}
}}\cr}$$
\medskip

\par
The construction of the  $E_{11}\otimes_s l_1$ non-linear realisation was  computed  in many of the early $E_{11}$ papers in various  dimensions for  the fields  at low levels,  but keeping only the usual coordinates of space-time and taking the local subgroup to be just the Lorentz group. However, the very unfamiliar nature of the many of the higher level  fields and especially  the coordinates inhibited the construction for a number of years.   Nonetheless, in 2007  all gauged maximal supergravities in five dimensions were  constructed by taking the fields to depend on some of the generalised coordinates [12]. In 2009 the papers    [17,18] computed the dynamics of the $E_{11}\otimes_s l_1$ non-linear realisation in four dimension keeping the 56 Lorentz scalar coordinates in addition to the usual four coordinates of space-time, but only  the fields at level zero, that is, the metric and the scalar fields. 
\par
Subsequently the $E_{11}\otimes_s l_1$ non-linear realisations was constructed in dimensions  four to seven  and 
corresponding  Lagrangians were constructed but keeping only the scalar fields and only the level one coordinates, that is, the coordinates of equation (1.2) [19]. This meant  discarding the graviton and  and even  the usual coordinates of space-time as well as all fields at all levels greater than zero, except the scalar fields,  and all coordinates at level greater than one [19]. In fact these  Lagrangians in
six and seven dimensions had previously been constructed [20], but the results of    [19] make it clear that they were just the non-linear realisation of 
$E_{11}\otimes_s l_1$ radically truncated in the way described. 
\par
A more systematic construction of the $E_{11}\otimes_s l_1$ non-linear realisation was  undertaken  more recently in eleven dimensions [21] and in four dimensions [22]. In the former paper the eleven-dimensional  equations of motion involving the fields at levels up to and including the dual graviton (level 3) and the generalised  coordinates $x^a$, $x_{a_1a_2}$ and $x_{a_1\ldots a_5}$ were given. While in reference [22], the four dimensional equations involving the scalars and vectors and the coordinates at levels zero and one, that is, the usual coordinates of space-time and the Lorentz scalar coordinates in the 56-dimensional representation of $E_7$,   were given. It was found that once one specified the Lorentz,  and  in the case of four dimensions also the SU(8),   character of the equations they were unique. Furthermore these equations  when restricted to the usual space-time and fields did indeed agree with those of maximal supergravity,  although the equation relating the usual graviton to the dual graviton required further study to be sure of its precise form.  
\par
The non-linear realisation of $E_{11}\otimes_s l_1$ when evaluated at low levels for the fields and coordinates  provides detailed equations of motion that involve the usual supergravity fields and derivatives with respect to the space-time coordinates  as well as some of the higher level coordinates. It also provides a geometry for the generalised space-time which has a specified tangent space equipped with a tangent group and a generalised vielbein which is determined and indeed easily computed at low levels. 
\par
The $E_{11}$ conjecture can also be stated as follows; the complete low energy effective action of strings and branes is the non-linear realisation of $E_{11}\otimes_s l_1$. It has been  thought that the complete low energy effective actions  were  the maximal supergravity theories, however, it has become clear that there are fields in the $E_{11}\otimes_s l_1$ non-linear realisation that are needed in the complete theory and these fields are not contained in the usual formulations of supergravity.  Hence if it is shown that the $E_{11}\otimes_s l_1$ non-linear realisation does contain the dynamics of the supergravity theories when one truncates  the higher level fields and coordinates then one has in effect provided overwhelming evidence for the conjecture. It may be hoped that the existence of an $E_{11}$ symmetry in the low energy effective action can provide some hints on what is the  theory that underlies strings and branes.  Clearly the $E_{11}$ proposal is more ambitious than to just reformulate the supergravity theories. 
\par
As we have mentioned the symmetries of the $E_{11}\otimes_s l_1$ non-linear realisation essentially determine the dynamics, but these symmetries do not include the familiar general coordinate and gauge transformations. However, the dynamics, at least in the most recent calculations of references [21,22] ,  do seem to possess these symmetries. In the early  papers on $E_{11}$ it was hoped that one could carry out a simultaneous non-linear realisation with the conformal group as was done in reference [23] for gravity and diffeomorphism transformations, however, this has not been possible to implement for  the full $E_{11}$ symmetry.  Recently, by assuming compatibility  with the $E_{11}$ symmetries,  a  simple formula for  the local transformations, that is, the diffeomorphisms and gauge transformations and their higher level analogues has been given [24]. One finds that the local transformations are labelled by the $l_1$ representation and so there is one local transformation for each coordinate in the generalised space-time. 
\par
A separate development that  originated from the realisation that strings could wrap around circles and as a result the string fields possessed  not only the usual space-time momenta but also  momenta associated with the wrapping of the circle. The resulting theories were found to be invariant under T duality. Corresponding to the doubling of momenta a corresponding doubling of coordinates was then  introduced into the first quantised string theory [25] in such a way as  to obtain a theory that was manifestly T duality invariant. In 1993  a manifestly T duality field theory that contained the massless NS-NS fields of the superstring which depended on  doubled coordinates [26,27] was found. In this work a section condition that restricted the dependence of the fields on the  doubled space-time was introduced and the field equations were shown to be a consequence of a generalised Riemann curvature condition.  
\par
Beginning in 2009  a field theory, called doubled field theory, was given. It  contained the massless NS-NS fields of the superstring which depended on  the  doubled coordinates and a had section condition [28,29,30,31]. It was eventually revealed that doubled field theory was equivalent  [32] to the theory introduced by Siegel in references [26,27]. As a result we refer to this theory as Siegel theory. A new development of Siegel theory was its extension to contain the massless fields of the R-R sector of the superstring [34]. This result was found by constructing  the $E_{11}\otimes_s l_1$ non-linear realisation in the decomposition  appropriate to the IIA theory in  ten dimensions. The dynamics at level zero  was just   Siegel theory [33], but the dynamics at level one contained the massless fields of the R-R sector of the superstring [34].  We note that this work required  no section condition.  
\par
A more recent development has been the construction of the so called exceptional field theory [35-39]. The starting point for this development was   that 

\item{-} the $E_n$ symmetries can be lifted to the eleven dimensional theory. [5]

\item{-} The usual space-time is generalised to include the coordinates given in equation (1.2) [8]. 

The fields content of the exceptional field theories are essentially the usual supergravity fields and so it has largely the same field content as the fields of the $E_{11}\otimes _s l_1$ non-linear realisation at low levels. Hence the  main ingredients of exceptional field theory  are just those found previously  in  the non-linear realisation of $E_{11}\otimes_s l_1$ at low levels. Indeed this latter construction was explained in detail to one of the subsequent authors of exceptional field theory [40]. 
\par
To understand why the reader, and particularly the younger researcher,  could be forgiven for thinking that the $E_{11}$ approach was only remotely connected to exceptional field theory it is useful to note that the  reference to the $E_{11}$ approach in the first paper on exceptional field theory [35] was \  "......(For more ambitious proposals see [12-[14]).....". 

We resist the temptation to further comment on the misleading account given in reference [35], but we note that it did not reference paper [22]  which gave the field  equations of  the gauge and scalar fields in  the $E_{11}\otimes_s l_1$ non-linear realisation in four dimensions, that is,  the dynamics of these field which depended on the generalised spacetime of equation (1.2).  
\par
One aspect in which exceptional field theory differs from the $E_{11}$ approach is that in the latter the equations of motion are derived from the symmetries of the non-linear realisation, however, in the former one uses gauge and diffeomorphism transformations which were suggested by  the notion of a generalised Lie derivative [41] in the enlarged space-time of equation (1.2) [8].  However,  diffeomorphism and generalised gauge transformations were formulated in  paper [24] within the context of the $E_{11}\otimes_s l_1$ non-linear realisation. Their form was in fact determined by the symmetries of the latter. 
\par
In this paper we show that exceptional field theory not only has the generic features of   the $E_{11}\otimes_s l_1$ non-linear realisation at low levels, but that it is contained within it if one makes a radical  truncations.  In section two we review the parts of the $E_{11}\otimes_s l_1$ non-linear realisation that we require. In section three we evaluate the  non-linear  gauge transformations of reference [24] for the low level fields using  a parameter that carries a tangent space index. In section four we give the generalised field strengths that 
appear in the field equations of the $E_{11}\otimes_s l_1$ non-linear realisation. While in section five we introduce a matter field  in the context of the $E_{11}\otimes_s l_1$ non-linear realisation, which belongs to the $l_1$ representation,  and construct its  covariant derivatives which are evaluated at low levels. In section six we transfer the results of sections three and five  to a world volume point of view. We find an  alternative formulation of the transformation of the vielbein  which is given by 
$$
E _A{}^\Sigma \delta E_\Pi {}^A= (D^{\underline \alpha})_\Pi{}^\Sigma  
(D^{\underline \alpha})_\Sigma {}^\Theta \partial_\Theta \Lambda ^\Sigma  + \Lambda^\Gamma \partial_\Gamma E_\Pi{}^A E_A{}^\Sigma 
$$
where the parameter $\Lambda ^\Sigma$  has  a world index, 
$ E_\Pi {}^A$  denotes the vielbein that   lives on the generalised spacetime and $(D^{\underline \alpha})_\Pi{}^\Sigma$ are the representation matrices of the $l_1$ representation. 
We also find the transformation of a  world volume matter field $ V^{\Pi}$ which belongs to the $l_1$ representation; the result is given by 
$$
\delta V^{\Pi} = -\,V^\Lambda\,\big(D^{\underline \alpha}\big)_\Lambda{}^\Pi\,\big(D_{\underline \alpha}\big)_\Sigma{}^\Xi\,\partial_\Xi\Lambda^\Sigma + \Lambda^\Sigma\,\partial_\Sigma V^\Pi + \mu\,V^\Pi\,\partial_\Sigma \Lambda^\Sigma.
$$
Evaluating the two results at low levels,   and brutally truncating,  we find   variations for the low level  fields that agree with those  found in exceptional field theory [35,36]. The same applies to the generalised field  strengths found in section four.

\medskip
{\bf 2 Review of the non-linear realisation of $E_{11}\otimes_s l_1$. }
\medskip
In view of the results of this paper it is conceivable that readers may,  and indeed perhaps should,  want to learn the $E_{11}$ approach. As such we take the liberty of briefly reviewing this approach. 
\par
It has been conjectured that the low energy effective action of strings and branes is the non-linear realisation of semi-direct product of $E_{11}$ 
with its first fundamental representation denoted $l_1$ [5,8].  The commutators of this algebra, which is denoted by $E_{11}\otimes_sl_1$, can be written in the form 
$$
[R^{\underline \alpha} , R^{\underline \beta} ]= f^{\underline \alpha \underline \beta}{}_{\underline \gamma} R^{\underline \gamma}, \quad
[R^{\underline \alpha} , l_A]= -(D^{\underline \alpha} )_A{}^B l_B
\eqno(2.1)$$
where $R^{\underline \alpha}$ are the generators of $E_{11}$ and $l_A$ are the generators belonging to the $l_1$ representation,  as is encoded in the second commutator. The Jacobi identities imply that the matrices $(D^{\underline \alpha} )_A{}^B$ are a representation of the $E_{11}$ algebra, 
$$[D^{\underline \alpha}, D^{\underline \beta }]= f^{\underline \alpha \underline \beta}{}_{\underline \gamma} D^{\underline \gamma}
\eqno(2.2)$$
We use the indices $\underline \alpha, \underline \beta, \ldots $ to label the generators of $E_{11}$. 
A review of the $E_{11}$ algebra and indeed Kac-Moody algebras in general can be found in the book of reference [42]. Indeed in this reference the reader can find the  $E_{11}\otimes_s l_1 $ algebra when decomposed into representations of SL(11) is given up to  level four. 
\par
The non-linear realisation of $E_{11}\otimes_sl_1$ is constructed from the group element $g\in E_{11}\otimes_sl_1$ that  can be written as 
$$
g=g_lg_E 
\eqno(2.3)$$
In this equation  $g_E$ is a group element of $E_{11}$ and so can be written in the form 
$g_E=e^{A_{\underline \alpha} R^{\underline \alpha}}$  and $A_{\underline\alpha}$ are the fields in the non-realisation. The group element   $g_l$ is formed from the generators of the $l_1$ representation and so has the form $e^{z^A l_A} $ where $z^A$ are the coordinates of the generalised space-time. The fields $A_{\underline\alpha}$ depend on the coordinates $z^A$.  The explicit form of these group elements  can be found in earlier papers on $E_{11}$,  such as in five dimensions in [12],  and more recently in  eleven dimensions in [21] and four dimensions in [22]. 

The non-linear realisation is, by definition, invariant under the transformations 
$$
g\to g_0 g, \ \ \ g_0\in E_{11}\otimes _s l_1,\ \ {\rm as \  well \  as} \ \ \ g\to gh, \ \ \ h\in
I_c(E_{11})
\eqno(2.4)$$
The group element $g_0\in E_{11}$ is a rigid transformation and so it is  a constant,  while the group element $h$ belongs to the Cartan involution subalgebra of $E_{11}$, denoted $I_c(E_{11})$; it is a local
transformation, that is,  it depends on the generalised space-time. As the generators in $g_l$ form a representation of $E_{11}$ the above transformations for $g_0\in E_{11}$ can be written as 
$$
g_l\to g_0 g_lg_0^{-1}, \quad g_E\to g_0 g_E\quad {\rm and } \quad g_E\to g_E h
\eqno(2.5)$$
\par
The Cartan involution   is an automorphism of the algebra, that is $I_c(AB)= I_c (A)I_c(B)$ for any two elements of the Lie algebra $A$ and $B$ and an involution, that is,  $I_c^2(A)= A$. It takes positive root generators to negative root generators and its action can be taken to be 
$$
I_c(R^{\underline \alpha}) = - R^{-\underline \alpha} 
\eqno(2.6)$$ 
for any root $\underline \alpha$. The Cartan Involution subalgebra is generated by 
$R^{\underline \alpha} - R^{-\underline \alpha} $. For the familiar finite dimensional semi-simple Lie groups the Cartan Involution invariant subalgebra is  the maximal compact of subalgebra; for example for SL(2,R) the Cartan Involution subalgebra is SO(2), while for $E_8$ it is SO(16). 
\par
Equation (2.6) leads to a theory that lives in a space with Euclidean signature. However, one can adopt an alternative Cartan Involution  by scattering some minus signs and one finds that the theory has a Minkowski signature [43]. As it is tiresome to keep track of these minus signs,  it is easier to work with  the Cartan Involution that leads to the Euclidean signature and then change to Minkowski signature at the end. This is the course of action we will take in this paper.  
\par
It follows from equation (2.5) that the coordinates are inert under the local transformations but transform under the rigid  transformations as 
$$
z^A l_A\to z^{A\prime} l_A=g_0 z^Al_A g_0^{-1} = z^\Pi D(g_0^{-1})_\Pi {}^Al_A
\eqno(2.7)$$
When  written  in matrix form the differential  transformations act as  
$dz^T
\to dz^{T\prime}= dz ^T D(g_0^{-1})$.   As a result the derivative
$\partial_\Pi\equiv {\partial\over \partial z^\Pi}$ in the generalised space-time 
transforms as $\partial_\Pi^\prime= D(g_0)_\Pi{}^\Lambda \partial_\Lambda$. 
\par 
The $l_1$ representation   of  $E_{11}$ is, by definition,  given by 
$$
U(k)( l_A)\equiv k^{-1} l_A k= D(k)_A{}^B l_B, \quad k\in E_{11}
\eqno(2.8)$$ 
where  $D(k)_A{}^B $  is the
matrix representative. For an infinitesimal transformation $k=1+a_{\underline \alpha} R^{\underline \alpha}$ and the matrix takes the form  
$D(k)_A{}^B= \delta_A^B + (D^{\underline \alpha})_A{}^B$. As a result, 
we recognise  the matrix $D(g_0^{-1})$, that appears in equation (2.7) as the matrix representation of the $l_1$ representation, although  the indices are labelled according to the role which they will play later in the physical theory that emerges from the non-linear realisation.  
\par
The dynamics of the non-linear realisation is just an action, or set of equations of motion, that are invariant under the transformations of equation (2.4). We  now recall how  to construct the dynamics of the 
the $E_{11}\otimes_s l_1$ non-linear realisation using the  Cartan forms  which are given by 
$$
{\cal V}\equiv g^{-1} d g= {\cal V}_E+{\cal V}_l, 
\eqno(2.9)$$
where 
$$
{\cal V}_E=g_E^{-1}dg_E\equiv dz^\Pi G_{\Pi, \underline \alpha} R^{\underline \alpha},
 \eqno(2.10)$$
belongs to the $E_{11}$ algebra and are the Cartan form for $E_{11}$, while  
 the part  in the space of generators of the $l_1$ representation is given by 
 $$
{\cal V}_l= g_E^{-1}(g_l^{-1}dg_l) g_E= g_E^{-1} dz\cdot l g_E\equiv 
dz^\Pi E_\Pi{}^A l_A  
\eqno(2.11)$$
\par
While  both ${\cal V}_E$ and ${\cal V}_l$ are invariant under rigid transformations, but  under local transformations of equation (2.5) they change as the 
$$ 
{\cal V}_E\to h^{-1}{\cal V}_E h + h^{-1} d h\quad {\rm and }\quad 
{\cal V}_l\to h^{-1}{\cal V}_l h 
\eqno(2.12)$$

\par
Although the 
Cartan form is inert under rigid transformations,  the rigid transformations do  act on
the coordinate differentials contained in the Cartan form and this action  induces  a corresponding  rigid $E_{11}$ transformation   on the lower index of ${ E_\Pi{}^A}$. On the other hand,   a local $I_c(E_{11})$ transformation acts on the $\underline \alpha$ index of $G_{\Pi, \underline \alpha}$ and on the $A$ index of ${E}_\Pi{}^{A} $ as governed by  equation (2.12).    We may summarise these two results as 
$$
{ E}_\Pi{}^{A\prime} =
D(g_0)_\Pi{}^\Lambda { E}_\Lambda{}^{B}D(h)_B{}^A \quad {\rm or}\quad \quad  (E^{-1})_A{}^{\Pi\prime}= D(h^{-1})_A{}^B (E^{-1})_B{}^\Lambda 
D(g_0^{-1})_\Lambda{}^\Pi
\eqno(2.13)$$
 \par
Thus   $E_\Pi{}^A$ transforms
under a local $I_c(E_{11})$ transformation on its $A$ index and by a rigid $E_{11}$ induced 
coordinate transformation of the generalised space-time on its $\Pi$ index.  These transformations mean that we can interpretation of ${ E}_\Pi{}^{A}$ as a generalised vielbein of the generalised space-time which possess a 
generalised tangent space that has  the tangent group $I_c(E_{11})$.  
\par
Examining equation (2.11) and recalling equation (2.8),  we recognise 
${ E}_\Pi{}^{A} $ as the representation  matrix $D(g_E)_\Pi{}^A$, and so 
${ E}_\Pi{}^A = D(g_E)_\Pi{}^A= (e^{A_{\underline \alpha}D^{\underline \alpha}})_\Pi{}^A$. It is useful to take the Cartan form to be in the 
 $l_1$ representation, taking the commutator of $G_{\Pi ,}{}_{\underline \alpha}
R^{\underline \alpha} $ of equation (2.10) with $l_A$ we find that 
$$
G_{\Pi ,}{}_A{}^B l_B\equiv G_{\Pi ,}{}_{\underline \alpha} (D^{\underline \alpha})_A{}^B l_B = (E^{-1})_A{}^\Lambda \partial_\Pi{} { E}_\Lambda{}^{B} l_B= - [G_\Pi{}_{\underline \alpha}
R^{\underline \alpha} ,l_A ] 
\eqno(2.14)$$
Converting  the first index on the Cartan form $G_{\Pi ,}{}_A{}^B$ to be a tangent index and using   the generalised vielbein  we can write The Cartan form in tangent space as 
$$
G_{A,}{}_B{}^C \equiv (E^{-1})_A{}^\Pi G_\Pi{}_B{}^C 
= (E^{-1})_A{}^\Pi(E^{-1})_B{}^\Lambda \partial _\Pi E_\Lambda{}^C 
\eqno(2.15)$$

Under the $I_C(E_{11})$ local transformation of equation (2.12) we find,  using equation (2.14),  that 
$$
G^{\prime}_{\Pi ,}{}_A{}^B = D(h^{-1})_A{}^C G_{\Pi ,C}{}^D D(h)_D{}^B
+D(h^{-1})_A{}^C  \partial_{\Pi } D(h)_C{}^B 
\eqno(2.16)$$
\par
 Thus if one chooses to construct the dynamics using the Cartan forms  $G_{A,}{}_B{}^C$ one just has to ensure that the equations of motion, or the action, are invariant under the local $I_C(E_{11})$  transformations. 
Furthermore the non-linear realisation automatically comes equipped with a generalised geometry in that it has a generalised tangent space, with tangent group $I_c(E_{11})$,   and a generalised 
vielbein $E_A{}^\Pi$.  
\par
As the dynamics is just that which is invariant under the transformation of equation (2.5), or equivalently (2.12),  it follows that this is, for a general non-linear realisation, determined by the group theory of the non-linear realisation. The worst that can happen is that there are  several constants whose values are not fixed,  but it is very often the case that the dynamics is uniquely fixed. 
\par
In this paper we will consider the $E_{11}\otimes_s l_1$ non-linear realisation in five dimensions. The algebra, fields, coordinates, generalised vielbein and all the data need for the construction are given in appendix A. Here, for convenience,  we record the fields which are given by 
$$
h_a{}^b, \ \varphi_{\alpha}, \ A_{aM}, \ A_{a_1a_2}{}^{N}, \ A_{a_1a_2a_3,\,\alpha}, \ A_{a_1a_2,\,b}, \ \ldots
\eqno(2.17)$$
and the coordinates of the generalised space-time which are 
$$
x^a, \ x_{N}, \ x_{a}{}^{N}, \ x_{a_1a_2,\alpha}, \ x_{ab}, \ \ldots
\eqno(2.18)$$


\medskip
{\bf 3 The gauge transformations  }
\medskip

The $E_{11}\otimes_s l_1$ non-linear realisation does not obviously possess diffeomorphisms and gauge transformations as one of its  symmetries.  However, it  was shown in reference  [24] how the $E_{11}$ symmetries of the non-linear realisation could be used to determine the transformation laws of these, and other,   local symmetries; the result was   given by 
$$
E^{-1}{}_A{}^\Pi \delta E_\Pi {}^B= (D^{\underline \alpha})_A{}^B (D_{\underline \alpha})_C{}^D D_D\Lambda ^C 
\eqno(3.1)$$
where $\Lambda ^C$ is the parameter of the gauge transformation,  which depends on the generalised space-time, and 
$D_D \Lambda^C= E_D{}^\Pi (\partial_\Pi \Lambda ^C- V^E\Omega _{\Pi , E }{}^C )$ 
and $\Omega _{\Pi , E }{}^C$ is the  connection which will be  discussed in  sections five and six. The repeated  $\underline \alpha$ index means a sum  over all such $E_{11}$ indices. The parameter $\Lambda^C$ carries the indices of the $l_1$ representation which leads to the generalised space-time. Thus there is a one to one correspondence between local transformations given in equation (3.1) and  the coordinates of space-time  as well as   the brane charges. Despite its unfamiliar appearance,  equation (3.1)  does contains the usual general coordinate and gauge transformations [24]. 
\par
At the linearised level the transformations of equation (3.1) reads [24]
$$
\delta {\cal A} _{\underline \alpha} = (D_{\underline \alpha})_D{}^C \partial_C \Lambda ^D , \quad {\rm or \ equivalently } \quad \delta {\cal A} _A{}^B  = (D^{\underline \alpha})_A{}^B(D_{\underline \alpha})_D{}^C \partial_C \Lambda ^D 
\eqno(3.2)$$ 
where at this level 
$E_\Pi {}^A= \delta _\Pi ^A + {\cal A} _{\underline \alpha} (D ^{\underline \alpha} )_\Pi {}^A = \delta _\Pi ^A + {\cal A}_\Pi {}^A$. These lowest level variations were evaluated at low levels in four, five and eleven dimensions in reference [24]. 
\par
We now evaluate the gauge transformations of equation (3.1) in five dimensions at low levels, but including the non-linear terms. 
The $E_{11}\otimes_S l_1$ algebra, the fields and coordinates of the non-linear realisation and generalised vielbein and the other quantities we need are reviewed in appendices  A and B.   Taking  $A=a$ and $B=N$ in equation (3.1) we find that the only non-vanishing  $(D^{\underline \alpha})_A{}^B$ is for $\underline \alpha = aN$. Using that  the Cartan-Killing metric is given by $g^ {aN}{}_{bM}= \delta ^a_b\delta _M^N $,  we find that 
$$
(D_{aN})_C{}^D = g^{-1} _{AN}{}_{\underline \alpha } (D^{\underline \alpha})_C{}^D = (D_{aN})_C{}^D 
\eqno(3.3)$$
Examining the  content of the $l_1$ representation of equation (A.3), and using that $(D^{aN})_{bM}= -\delta ^a_b\delta_M^N$,  we find that equation (3.1), for our chosen values of $A$ and $B$,  is given by 
$$
 E^{-1}{}_a{}^\Pi \delta E_\Pi {}_N
= -(D_{aN})_D{}^C D_C \Lambda ^D
$$
$$
= -(D_{aN})^P{}_{\Vert}{}^b D_b \Lambda_P- (D_{aN})^b{}_P{}_{\Vert}{}_Q D^Q \Lambda^R_b
- (D_{aN})^{c_1c_2\beta}{}_{\Vert}{}_b^Q D^b{}_Q \Lambda_{c_1c_2\beta}+\ldots
$$
$$
= D_a \Lambda_N - 10\,d_{NMP}\,D^M \Lambda_a{}^P - 12\,\left(D^\alpha\right)_N{}^M\,D^b{}_M \Lambda_{ab,\,\alpha} + \ldots 
\eqno(3.4)$$ 
Care has to be taken with the position of the indices, for example 
taking $B=N$ results in an $N$ index which is downstairs,  as it should be when it appears on the vielbein, rather than the inverse vielbein. Looking at the definition of the vielbein of equation (2.11) we indeed see that it should be in the opposite position to the $N$ index on  $ Z^N$. Given the subtleties with the index placements we will often use a $\Vert$ to make clear the separation of the $A$ and $B$ indices on $D^{\underline \alpha}$ and the connection. 
\par
We now evaluate the left-hand side of equation (3.4). Looking at the five dimensional vielbein of equation (A.28) and its inverse of equation (A.32) we find that it is equal to 
$$
E^{-1}{}_{a \Vert}{}^\Pi \delta E_\Pi {}_{\Vert N}= E^{-1}{}_{ \Vert a}{}^\nu \delta E_\nu{ }_{\Vert N}+E^{-1}{}_{a\Vert }{}_{\dot M}  \delta E^{\dot M} { }_{\Vert N}
+E^{-1}{}_{a\Vert }{}_\nu^{\dot M} \delta E^\nu_{\dot M} { }_{\Vert N}
$$
$$
= -e^{-1}_a{}^\nu \delta (e_\nu{}^b A_{bN})+ A_{aP} d^P_{\dot M} \delta (d_N{}^{\dot M})= -e^{-1}_a{}^\nu \delta  A_{b\dot M}  (d_N{}^{\dot M})
\eqno(3.5)$$
We note that the vielbein has a universal factor of $ (\det  e)^{-{1\over 2}}$ but this,  and its derivative, drop out of the above expression . 
Taking  this equation together with the right-hand side of  equation (3.4) we find  that 
$$
\delta A_{\mu \dot N}=-e_\mu {}^a d_{\dot N} ^M (D_a \Lambda_M - 10\,d_{MRP}\,D^R\Lambda_a{}^P - 12\,\left(D^\alpha\right)_M{}^P\,D^b{}_P \Lambda_{ab,\,\alpha} + \ldots )
\eqno(3.6)$$
\par
We now consider the right-hand side of equation (3.1) when $A=a_1$ and $B= {}^N{}_{a_2}$ whereupon  it takes the form 
$$
E_{a_1\Vert}{}^\Pi\,\delta E_{\Pi \Vert a_2}{}^N = (D^{\underline \alpha})_{a_1\Vert a_2}{}^N (D^{\underline \alpha})_C{}^D D_D\Lambda ^C
$$
$$
={1\over 10} (D_{a_1a_2}{}^N)_C{}^D D_D\Lambda ^C= 
{1\over 10}\{(D_{a_1a_2}{}^N)^{b_1}{}_{M\Vert}{}^{b_2}\,D_{b_2} \Lambda_{b_1}{}^M 
$$
$$+ \,\left(D_{a_1a_2}{}^N\right)^{b_1b_2,\,\alpha}{}_{\Vert M}\,D^M \Lambda_{b_1b_2,\,\alpha} + \,\left(D_{a_1a_2}{}^N\right)^{b_1b_2}{}_{\Vert M}\,D^M \Lambda_{b_1b_2} + \ldots\}
$$
$$
= 2 D_{[a_1}\,\Lambda_{a_2]}{}^N -12\left(D^\alpha\right)_M{}^N D^M \Lambda_{a_1a_2,\,\alpha} +{2\over 3} \left(D^\alpha\right)_M{}^N D^M \Lambda_{a_1a_2}
+ \ldots
\eqno(3.7)$$ 
\par 
To find the variation of the field we must evaluate the left-hand side of this equation; 
$$
E_{a_1\Vert}{}^\Pi\,\delta E_{\Pi \Vert a_2}{}^N = 
-2\delta (A_{\mu\nu }{}^{\dot M}) d_{\dot M}^N e_a{}^\mu e_b{}^\nu+ d^{\dot N \dot R\dot S} d_{\dot N}^N A_{[\mu | \dot R} \delta A_{|\nu ]\dot S}e_a{}^\mu e_b{}^\nu
\eqno(3.8) $$
In deriving this equation we have used the fact that $d^{NMP}$ is an $E_6$ invariant tensor and that $d_P^{\dot M}$ can be identified with an $E_6$ group element in the relevant representation. Taking these results for the left and right-hand sides of equation (3.1),  for our chosen indices,   we find that
$$
\delta A_{\mu \nu}{}^{\dot M}= e_\mu {}^a e_\nu {}^b d_P^{\dot M} (-D_{[a} \Lambda _{b]}{}^P +6 \left(D^\beta\right)_R{}^P D^R \Lambda_{a_1a_2,\,\beta} -{1\over 3} \left(D^\alpha\right)_R{}^P D^R \Lambda_{a_1a_2}+\ldots )
$$
$$
+{1\over 2} e_\mu {}^a e_\nu {}^b d^{\dot N \dot R\dot S} d_{\dot N}^N A_{[\mu | \dot R} \delta A_{|\nu ]\dot S}
\eqno(3.9)$$
\par

Finally, in equation (3.1), we take the values  $A=a_1$ and $B= {}_{a_1a_2\alpha}$ to find that the right-hand side of equation (3.1) reads 
$$
 -\,{3\over 360}\,\left(D_{a_1a_2a_3}{}^\alpha\right)^{b_1b_2,\,\beta\Vert b_3}\,D_{b_3} \Lambda_{b_1b_2,\,\beta} = -3D_{[a_1}\,\Lambda_{a_2a_3]}{}^\alpha  + \ldots
\eqno(3.10)$$
while the left-hand side is given by 
$$
E_{a_1 \Vert}{}^\Pi\,\delta E_{\Pi \Vert a_2a_3}{}^\alpha =-3 e_{a_1}{}^\mu \delta (e_\mu{}^b A_{ba_1a_2})+\dots $$
and as a result 
$$
\delta A_{\mu_1\mu_2\mu_3}= - D_{[\mu_1} \Lambda _{\mu_2\mu_3]\alpha}+\ldots 
\eqno(3.11)$$
\par
We will now evaluate the above gauge transformations at very low levels. To do this we will need to evaluate the covariant derivatives acting on the gauge parameters $\Lambda ^A$ which belong to  the $l_1$ representation.  In section five, with the help of  appendix B,  the covariant derivatives of the an object $V^A$ which transforms as the $l_1$ representation does under $I_c(E_{11})$ are introduced and evaluated at low levels. The covariant derivatives that appear in the gauge transformations have the same form as the the covariant derivatives acting on $V^A$ and the connections  act on the parameters $\Lambda^A$ in the same way. Nonetheless the detailed expressions for the connections may not be the same in the two different applications.  \par
We may write the 
gauge transformation of the field $A_{aN}$ as 
$$
\delta A_{\mu \dot N}=-e_\mu {}^a d_{\dot N} ^M {\cal D}_a \Lambda_M 
=e_\mu {}^a d_{\dot N} ^M (D_{aM})_C{}^D D_D\Lambda ^C
\eqno(3.12)$$ 
In section five this covariant derivative on $V^A$  was evaluated at low levels and in equation (5.14) truncated to contain only the component $V_N$, which corresponds here to  the component $\Lambda_N$, using this result we find that   
$$
\delta A_{\mu \dot N}
= -(\det e)^{{1\over 2}}d_{\dot N} ^N   \{\partial_\mu \Lambda_N +
\Lambda_P \Omega_{\mu ,} {}_{\alpha} (D^\alpha -D^{-\alpha})_N{}^P 
$$
$$
+A_{\mu P} d^P_{\dot P} (\partial^{\dot P}  \Lambda_N + 
\Lambda_M \Omega^{\dot P}{}_{ ,} {}_{\alpha} (D^\alpha -D^{-\alpha})_N{}^M  )
$$
$$
+10 d_{NMP} d_{\dot R}^M (\partial^{\dot R}\Lambda_a{}^P 
+\Lambda_L\Omega ^{\dot R} {}_{, \mu Q}d^{QLP} 
 +\ldots \}
\eqno(3.13)$$
where we have also keep one term which contains the level two parameter $\Lambda_a{}^P $ but discarded all other such terms. 
\par
Using the covariant derivative introduced in section five we can write the gauge variation of the  field $A_{\mu_1\mu_2}{}^{\dot M}$,  given in equation (3.9), in the form 
$$
\delta A_{\mu_1\mu_2}{}^{\dot M}= -{1\over 2}e_{\mu_1} {}^a e_{\mu_2} {}^b d_p^{\dot M} {\cal D}_{[a} \Lambda _{b]}{}^P 
+{1\over 2} d^{\dot N \dot R\dot S} A_{[\mu_1 | \dot R} \delta A_{|\mu_2 ]\dot S}
\eqno(3.14)$$
This can be further evaluated using the formulae in section five, especially equation (5.17).  
\par
We close this section by finding, at low levels, the gauge transformations of the level zero fields, namely the vielbein and the scalars. To find the variation of the vielbein $e_\mu{}^a$ we take $A=a$ and $B = b$ in equation (3.1) whereupon, using the vielbein and generalised vielbein of Appendix A, that is, equations (A.28) and (A.32),   we find that the left-hand side of this equation becomes
$$
E_a{}^\Pi\,\delta E_\Pi{}^b = (\det e)^{{1\over 2}}e_a{}^\mu\,\delta ((\det e)^{-{1\over 2}} e_\mu{}^b )= \,e_a{}^\mu \delta e_\mu{}^b - {1\over 2}\,\delta_a^b e_c{}^\lambda\,\delta e_\lambda{}^c
\eqno(3.15)$$
While the right-hand side of this equation is given by 
$$ E_a{}^\Pi\,\delta E_\Pi{}^b ={\cal D}_a \Lambda^b
= D_a\,\Lambda^b - D^b{}_N\,\Lambda_a{}^N - {1\over 2}\,\delta_a^b\,\left(D_c\,\Lambda^c + D^N\,\Lambda_N + D^c{}_N\,\Lambda_c{}^N\right)+\ldots 
\eqno(3.16)$$
where we have used equation (B.14). As a result we conclude that 
$$
e_a{}^\mu \delta e_\mu{}^b = D_a\,\Lambda^b - D^b{}_N\,\Lambda_a{}^N +\delta_a^b ({1\over 3}D^N\,\Lambda_N +{2\over 3}  D^c{}_N\,\Lambda_c{}^N )+\ldots .
\eqno(3.17)$$
Keeping only terms involving the field   $e_\mu{}^a$ and the parameter $\Lambda^a$ we recognise that this field has the standard transformation under general coordinate transformations. This justifies the choice of  $e_\mu{}^a= (e^h)_\mu{}^a$  to be the  correct  vielbein rather than the combination $\tilde e_\mu{}^a= (\det e)^{-{1\over 2}}e_\mu{}^a$ that appears in  the generalised vielbein given in  equation (A.28). 
\par
To find the variation of the scalars, that are contained in their vielbein $\delta d_N{}^{\dot N}$ we choose $A = N$ and $B=M$ in equation (3.1).  Using the vielbein and inverse vielbein given in  equations (A.28) and (A.32),  we find that the left-hand side of equation (3.1) is given by 
$$
E^{N\Pi}\,\delta E_{\Pi M} = (\det  e)^{{1\over 2}}d_{\dot N}{}^N\,\delta ((\det  e)^{-{1\over 2}})d_M{}^{\dot N} )= d_{\dot N}{}^N\,\delta d_M{}^{\dot N}-{1\over 2} \delta _M^N e_c{}^\lambda\,\delta e_\lambda{}^c\
\eqno(3.18)$$
  While, using equation (B.13)  the righthand side of equation (3.1) is given by 
$$
E^{N\Pi}\,\delta E_{\Pi M}= {\cal D}^N \Lambda_M = \left(D^{\underline \alpha}\right)^N {}_M\,\left(D_{\underline \alpha}\right)_A{}^B\,D_B\,\Lambda^A 
$$
$$
= { 6}\,\left(D^{\alpha}\right)_M{}^N\,\left(D_{\alpha}\right)_P{}^Q\,D^P\,\Lambda_Q - { 6}\,\left(D^{\alpha}\right)_M{}^N\,\left(D_{\alpha}\right)_P{}^Q\,D^a{}_Q\,\Lambda_a{}^P
$$
$$
+\,\left(D^a{}_b\right)^N{}_M\,\left(D_a{}^b\right)_c{}^d\,D_d\,\Lambda^c + \left(D^a{}_b\right)^N{}_M\,\left(D_a{}^b\right)^P{}_Q\,D^Q\,\Lambda_P + \left(D^a{}_b\right)^N{}_M\,\left(D_a{}^b\right)_c{}^{P,\,d}{}_Q\,D^d{}_Q\,\Lambda_c{}^P 
$$
$$
-\,{1\over 3}\left(\left(D^a{}_a\right)^N{}_M\,\left(D^b{}_b\right)_c{}^d\,D_d\,\Lambda^c + \left(D^a{}_a\right)^N{}_M\,\left(D^b{}_b\right)^P{}_Q\,D^Q\,\Lambda_P + \left(D^a{}_a\right)^N{}_M\,(D^b{}_b)_c{}^{P,\,d}{}_Q\,D^d{}_Q\,\Lambda_c{}^P\right) 
$$
$$
= D^N\,\Lambda_M - D^a{}_M\,\Lambda_a{}^N  - 10\,d^{NPR}\,d_{MQR}\,\left(D^Q\,\Lambda_P - D^a{}_P\,\Lambda_a{}^Q\right)
$$
$$
-{1\over 2} \delta_M^N(D_c\Lambda^c+D^p\,\Lambda_P +3 D^c{}_P\,\Lambda_c{}^P )+ \ldots\,.
\eqno(3.19)$$
In this last equation   we have used the expression of the Cartan-Killing metric of equation (B.13) and the identity of equation (A.14).  The variation of $d_N{}^{\dot N}$ can be found by using equation (3.18) and  equation (3.19)  to be given by  
$$
d^N{}_{\dot P} \delta d^{\dot P}_M = D^N\,\Lambda_M - D^c{}_M\,\Lambda_c{}^N- 10d^{MPR}\,d_{NQR}\,\left(D^Q\,\Lambda_P - D^c{}_P\,\Lambda_c{}^Q\right) 
$$
$$
+{1\over 3} \delta_M^N D^p\,\Lambda_P -{1\over 3}
\delta_M^N  D^c{}_P\,\Lambda_c{}^P 
$$
$$
= { 6}\,\left(D^{\alpha}\right)_M{}^N\,\left(D_{\alpha}\right)_P{}^Q\,D^P\,\Lambda_Q - { 6}\,\left(D^{\alpha}\right)_M{}^N\,\left(D_{\alpha}\right)_P{}^Q\,D^a{}_Q\,\Lambda_a{}^P+\ldots 
\eqno(3.20)$$
\par
In section six we compute the gauge variations of the fields  with a parameter that is a world vector rather than a tangent vector.  After a radical truncation we make contact with the analogous quantities found  in exceptional field theory. 
\par
We close this section by noting the relation between the gauge transformations of equation (3.1) and local $I_c(E_{11})$ transformations. We can write equation (3.1) in the form 
$$
E^{-1}{}_A{}^\Pi \delta E_\Pi {}^B= {1\over 2} (D^{\underline \alpha}+D^{-\underline \alpha})_A{}^B (D_{\underline \alpha})_C{}^D D_D\Lambda ^C 
+{1\over 2} (D^{\underline \alpha}-D^{-\underline \alpha})_A{}^B (D_{\underline \alpha})_C{}^D D_D\Lambda ^C 
\eqno(3.21)$$
We recognise the last term as an $I_c(E_{11})$ transformation of $E_\Pi{}^A$ 
and this we can discard from the gauge transformation as it is symmetry of the theory. We also note that the gauge transformations also contain the rigid $E_{11}$ transformations [24].


\medskip
{\bf 4 Construction of the generalised field strengths.  }
\medskip
As explained in section two,  if one uses  the Cartan forms to construct the dynamics then the invariance  under the rigid $E_{11}$ symmetry of the non-linear realisation is automatically encoded  and one just has to ensure invariance under the local subalgebra $I_C(E_{11})$.  This has been the strategy adopted in most of the previous $E_{11}$ papers. Indeed,  in [21] and [22]  a systematic effort was made to find the equations of motion of the $E_{11}\otimes l_1$ non-linear realisation  in  eleven and four dimensions respectively. The dynamics consists of a series of duality relations that are first order in derivatives with respect to the generalised coordinates and, at least at low levels, the dynamics   appears to be uniquely determined by the symmetries of the non-linear realisation. As such the dynamics consists of equations which were linear in the Cartan forms which, we recall from section two,  have the form $G_{A,\underline \alpha}$ where the first index $A$ arises from the $l_1$ part of the non-linear realisation and the 
second index $\underline \alpha$ is from the $E_{11}$ part. At the linearised level $G_{A,\underline \alpha}=\delta _A^\Pi \partial_{\Pi}A_{\underline \alpha}$ and as such if $A=a$ we get a lead term in the Cartan form that contains the  usual space-time derivative of a field while if $A$ is a higher level index we get a derivative with respect to one of the higher level coordinates. 
\par
Examining the resulting equations of motion of references [21] and [22],  it becomes apparent that the Cartan forms that contain a leading  term that has a usual space-time derivative, that is, $G_{a , \underline \alpha}$  came together with 
a Cartan form  whose leading term  contains the derivative with respect to one of the higher level generalised coordinates. 
The reader may look at equations (4.10), (4.11) and (4.14-17) of reference [22] to see how this works in four dimensions using just the symmetries of the $E_{11}\otimes_s l_1$ non-linear realisation. 
\par
In reference [24] the same phenomenon was observe in four and five dimensions,  
but instead using the rigid $E_{11}$ symmetry instead of the local $I_c(E_{11})$ symmetry. Since we are considering five dimensions in this paper, we now recall  how this goes in this dimension. The equation of motion of the field $A_{aN}$ must contain the Cartan form $G_{[a_1 , a_2] N}$ as at the linearlised order it is  of the form 
$G_{[a_1 , a_2] N}\sim \partial _{ [a_1} A_{a_2] N}$. It was found  that  such  a term in the equation of motion occurs in the combination  
$$
\partial _{[a_1 }A_{a_2] N}+ 10 d_{NRS} \partial^R A_{a_1a_2}{}^{S}
\eqno(4.1)$$
In terms of Cartan forms this means that the equations of motion involve the combination 
$$
{\cal G} _{a_1 a_2 N}= G_{[a_1 ,a_2] N}+ 10 d_{NRS} G^R {}_{, a_1a_2}{}^{S}
\eqno(4.2)$$
\par
The equation of motion of the field $A_{aN}$ implies that its field strength, which is contained in $G_{[a_1, a_2] N}$, is dual to the field strength of the field $A_{a_1a_2}{}^{ N}$ 
which is continued in the Cartan form $G_{[a_1 , a_2 a_3 ]}{}^N$. At the linearised level the latter is of the form 
$G_{[a_1 , a_2 a_3 ]}{}^N\sim \partial _{[a_1 } A_{ a_2 a_3 ]}{}^N$

Carrying out the same procedure for the field $A_{a_1a_2}{}^{S}$ we find  that it occurs in the linearised combination 
$$
{\partial}_{[a_1}A_{  a_2a_3]}{}^N +6 (D^\beta )_R{}^N 
\partial ^R A_{, a_1a_2a_3 \beta} 
\eqno(4.3)$$
and as a result the associated Cartan forms must occur in the combination  
$$
{\cal G}_{a_1  a_2a_3}{}^{S}= G_{[a_1 , a_2 a_3 ]}{}^N +6 (D^\beta )_R{}^N 
G^R{}_{, a_1a_2a_3 \beta} 
\eqno(4.4)$$
We refer to the modified field strengths, such as in equations (4.2) and (4.4), as generalised field strengths. 
Consequently, the equation of motion for the vector field should relate the generalised field strength of equation (4.2) to the dual of the generalised field strength of equation (4.4). 
\par 
An alternative  way to find these combinations is to use the gauge transformations of equation (3.1). It suffices to do this using the linearise expressions for the fields in the  Cartan forms and their lowest order  gauge transformations. One finds that ${\cal G} _{[a_1 }A_{a_2] N}$ and ${\cal G}_{a_1  a_2a_3}{}^{S}$ are invariant under the lowest level gauge transformations of equation (3.1), that is, those given in equation (3.2). These latter transformations for the lowest level fields are  [24] 
$$
\delta A_{aM}= -\partial_a \Lambda_M + 10 d_{MPQ} \partial^Q\Lambda^Q_a + 12 (D^\beta)_M{}^N\partial_N^b\Lambda_{ba\beta}+\dots 
$$
$$
\delta A_{a_1a_2}{}^{M}=-\partial_{[a_1} \Lambda_{a_2]}^M +6 (D^\beta)_N{}^M\partial^N\Lambda_{ba_1a_2\beta}+\dots 
$$
$$
 \delta h_a{}^b= \partial_a \xi^b+ {1\over 3}  \delta _a^b\partial ^M\Lambda _M - \partial_N^a \Lambda_b{}^N+{2\over 3} \delta_a^b \partial_N^c \Lambda_c{}^N+\ldots , 
$$
$$
 \delta \phi^\alpha = -6 (D^\alpha)_P{}^R \partial^P\Lambda_R
+6 (D^\alpha)_P{}^R \partial_R^b\Lambda_b{}^P +\dots
\eqno(4.5)$$
The last equation can also be written as 
$$
 \delta \phi_M{}^N = 6 (D^\alpha)_M{}^N (D^\alpha)_P{}^R (\partial^P\Lambda_R
- \partial_R^b\Lambda_b{}^P) +\dots
$$
where $\delta \phi_M{}^N =  -(D^\alpha)_M{}^N \phi_\alpha$. It is instructive to  recall that the scalar vielbein is defined  by the equation 
$e^{-\phi_\alpha R^\alpha} Z^{\dot N} e^{\phi_\alpha R^\alpha}= Z^M (d^{-1})_M{}^{\dot N}$. One then finds that $(d^{-1})_{ M}{}^{\dot N}= (\exp (-D^\alpha \phi_\alpha))_{ M}{}^{\dot N} $. 
\par
Applying these arguments to the scalars we find that the quantity 
$$ \partial_a \phi_\alpha - 6 (D^\alpha)_P{}^R \partial^P A_{a R}
\eqno(4.6)$$
is gauge invariant at lowest level under $\Lambda _N $ transformations and so the non-linear quantity that should appear in the equations of motion is given by 
$$
{\cal G}_{a, \alpha} = G_{a, \alpha} -6  (D_\alpha)_P{}^R G^P{}_{, a R}
\eqno(4.7)$$
For the graviton it is the quantity 
$$
\partial _a h_b{}^c + {1\over 3} \delta _b^c \partial ^P A_{aP} 
\eqno(4.8)$$
which is invariant under under lowest order $\Lambda _N $ transformations and this in turn implies one should in the equations of motion use the quantity 
$$
{\cal G}_a{}_{, b}{}^c = G_a{}_{, b}{}^c + {1\over 3} \delta _b^c G^P {}_{,aP}
\eqno(4.9)$$
Of course for the scalars one must also take into account the standard  local symmetry of the non-linear realisation $E_6$ with local subgroup USp(8) which requires one to use the part of the Cartan form that is in the coset and so transforms homogeneously. While for the vielbein we must also ensure general invariance. In fact the unique object composed from the vielbein and  which has only one derivative that  is invariant under general coordinate transformations  at lowest order is the spin connection. The above discussion suggests that this should be modified by replacing derivative of the vielbein by the combination of equation (4.9). 
\par
We now give explicit expressions for the objects in equations (4.2) and (4.4). The Cartan forms 
$G_{\Pi , \underline \alpha}$ are defined in equation (2.11). They are very simple to evaluate using their definition in equation (2.10) and they have been extensively computed in previous $E_{11}$ papers,   for example in five dimensions in the  [12,44]. They are given by 
$$
G_{\Pi ,a N}= \partial_\Pi A_{aN}
\eqno(4.10)$$
$$
G_{\Pi , a_1 a_2}{}^{ N}= \partial_{\Pi} A_{a_1a_2}{}^{ N}+{1\over 2} d^{NPQ}  \partial_\Pi A_{[ a_1 | P} A_{| a_2 ] Q}
\eqno(4.11)$$
$$
G_{\Pi , a_1a_2a_3 \beta}= \partial_\Pi A_{ a_1 a_2 a_3 \beta}-{1\over 6} \partial_\Pi A_{[ a_1 | M} A_{| a_2 | N} A_{| a_3 ]P} d^{MNQ} (D^\alpha )_Q{}^P 
-\partial_\Pi A_{[ a_1a_2}{}^M A_{a_3 ]N} (D^\alpha )_M{}^N 
\eqno(4.12)$$
\par 
The Cartan forms of equations (4.2) and (4.4) have as their first index a tangent index rather than the world index  that they possess in equations (4.10-12). The quantities are related by $G_{A, \underline \alpha}= E^{-1}{}_A {}^\Pi G_{\Pi, \underline \alpha}$ where the inverse vielbein $E^{-1}{}_A {}^\Pi $ is given in appendix A in equation (A.32). Evaluating the effect of the inverse vielbein we find for example that 
$$
G_{a, \underline \alpha} = e_a{}^\mu G_{\mu , \underline \alpha} + A_{aM}d^M_{\dot N}  G^{\dot N}{}_{, \underline \alpha} + (2A_{a \nu}{}^M d_M^{\dot P} -{1\over 2} d_M^{\dot P} d^{MRS} A_{a R} A_{\nu S} ) G^\nu{}_{\dot P , \underline \alpha}+\ldots 
\eqno(4.13)$$
\par
We now truncate the quantities  of equations (4.2) and (4.4),  keeping only the fields $A_{aN}$ and $A_{a_1a_2 }{}^N$ and  the derivative $\partial_\mu$ and $\partial^{\dot N}$,  and discarding all higher level terms to find the expressions 
$$
{\cal G} _{a_1 a_2 N}= e_{[a_1}{}^\mu \partial_\mu A_{\mu a_2] N}+ A_{[a_1 \dot N}\partial^{\dot N} A_{a_2] Q}+ 
10 d_{NRS} d_{\dot R}^R( \partial^{\dot R} A_ {a_1a_2}{}^{S}+ {1\over 2}d^{SPQ}\partial^{\dot R} A_{[ a_1  | P} A_{|a_2 ] Q})
\eqno(4.14)$$
and 
$$
{\cal G}_{a_1  a_2a_3}{}^{N}= e_{ [a_1}{}^\mu (\partial_\mu A_{a_2a_3 ]}{}^{ N}+{1\over 2} d^{NPQ}  \partial_\mu A_{| a_2 | P} A_{| a_3 ]Q})
+ A_{[a_1 | P} d^P_{\dot P }\partial^{\dot P} A_{ | a_2a_3]}{}^{ N}
$$
$$
+{1\over 3} d^{NRQ} d^P_{\dot P } A_{[a_1 | P}\partial^{\dot P} A_{| a_2 |R} A_{|a_3 ] Q}
-{5\over 3} d^{NQW} d^{MPL} d_{\dot R}^R\partial^{\dot R} A_{[a_1 | M} A_{| a_2 | P} A_{| a_3] Q}d_{LRW} 
$$
$$
+ A_{[ a_1a_2 }{}^N d_{\dot T}^S\partial ^{\dot T} A_{a_3 ] S}
-{1\over 3} d_{\dot T}{}^N \partial ^{\dot T} A_{[a_1a_2 }{}^{S}A_{a_3] S} 
-10 d^{NTW} d_{RSW} d_ {\dot M} ^ R A_{[a_1a_2 }{}^{S}\partial^{\dot M}A_{a_3] T} 
\eqno(4.15)$$
In deriving this result we have used the identity of equation (A.14) and we have redefined $A_{a_1a_2a_3\alpha}$ before discarding it.  After a suitable field redefinition the generalised field strengths of equation (4.14) and (4.15) agree with the generalised field strengths given in the papers  [35,36] on exceptional field theory in five dimensions. The possible exception is the coefficient of one term whose value in the papers on exceptional field theory we are not sure of. We will assume that this coefficient does in fact agree on closer examination. 
\par
The equation of motion for the vector was deduced using the symmetries of the non-linear realisation $E_{11}\otimes_s l_1$ in four dimensions [22]. It essentially stated that the generalised field strength were self-dual. It is inevitable that a similar calculation in five dimensions will lead to a  condition self-duality condition relating the generalised field strength in equation (4.14) and (4.15).

\medskip
{\bf 5 Matter representations }
\medskip
Matter representations for non-linear realisations were introduced in the classic papers of reference [45] for the case when the group did not include space-time generators. These are objects which transform linearly under the local subgroup.  In previous papers on $E_{11}$ such matter representations were not discussed as they are not needed for the development of the theory, but it is straight forward to introduce them. We recall that the generators of the $l_1$ representation by definition belong to a representation of $E_{11}$, a fact that is  encoded in the second commutator of equation (2.1),  and they transform under a finite $E_{11}$ transformation as in equation (2.7). As a result they must transform under the local subgroup, $I_c(E_{11})$ as follows 
$$
h^{-1} l_A h= D(h)_A{}^B l_B 
\eqno(5.1)$$
where $h\in I_c(E_{11})$. Such an infinitesimal transformation is of the form $h= 1+a_{\underline \alpha}(R^{\underline \alpha}- R^{-\underline \alpha})$.
\par
We can introduce a $l_1$  matter representation $V^A$ whose transformation is such that $V^{A\prime}  l_A= h^{-1}V^A l_A h$ and as a result it transforms as 
$$
V^{A\prime}= V^B D(h)_B{}^A= V^A+ V^B (D^{\underline \alpha}- D^{-\underline \alpha}) _B{}^A a_{\underline \alpha}+ \dots 
\eqno(5.2)$$
We observe that this is the same way as the $A$ index on the generalised vielbein $E_\Pi{}^A$ transforms. Examining the components of the generators of the $l_1$ representation given in appendix A we find that the representation $V^A$ has the components 
$$
V^A= \{ V^a , V_N, V_a^N, V_{c_1c_2\alpha} , V_{c_1c_2}, \ldots \}
\eqno(5.3)$$
\par
We define the covariant derivative of $V^A$ to be 
$$
{\cal D}_A V^B = (D^{\underline \alpha} )_A{}^B (D_{\underline \alpha} )_C{}^D D_D V^C
\eqno(5.4)$$
where the sum is over all indices $\underline \alpha$. 
In this equation we  define
$$
D_{\underline \alpha} = g_{\underline \alpha \beta } D^{\underline \beta}
\eqno(5.5)$$
where  
$ g_{\underline \alpha \beta }$ is the inverse of $ g^{\underline \alpha \beta }$, which is the Cartan-Killing metric of $E_{11}$ which is given in appendix A. We also take 
$$
D_D V^C= E_D{}^\Pi (\partial _\Pi V^C- V^D\Omega _{\Pi ,}{}_D{}^C )
\eqno(5.6)$$
Where $\Omega _{\Pi ,}{}_D{}^C$ is the  connection which changes  under a local $I_c(E_{11})$ transformation  as          
$$
\Omega _{\Pi }{}^{\prime}= D(h^{-1} )\Omega _{\Pi}D(h) + D(h^{-1} )\partial _\Pi D(h) 
\eqno(5.7)$$
where in this last equation we have viewed $\Omega _{\Pi ,}{}_D{}^C$ as a matrix and $D(h)$ is the matrix $D(h)_A{}^B$ for $h\in I_c(E_{11})$.  
The connection is to be constructed from the Cartan forms of equation (2.9). Examining equation (2.16) we recognise that these Cartan forms have the correct transformation law of equation (5.7). In fact the part of the Cartan form that belongs to  the local subalgebra has precisely  this transform and so from this viewpoint alone one could  just take this to be the connection. However, it will turn out that the connection is the $I_c(E_{11})$ part of the Cartan form  plus a specific combination of the Cartan forms that belong to the coset part which transform covariantly under  $I_C(E_{11})$. This is apparent when one considers general relativity from this viewpoint. It is usual to take the connection, no matter how it is constructed, to belong in the local subalgebra which would imply that 
$\Omega _{\Pi ,}{}_D{}^C= 
\Omega_{\Pi ,\underline \alpha }(D^{\underline \alpha}-D^{-\underline \alpha})_D{}^C$, however, in this paper we will take a more liberal perspective. 
\par 
The derivative  $D_D V^C$  transforms covariantly under a local $I_C(E_{11})$ as one might expect 
$$
(D_D V^C)^\prime = D(h^{-1})_D{}^E D_E V^F D(h)_F{}^C
\eqno(5.8)$$
This is not an irreducible representation of $I_c(E_{11})$. We can project it onto the adjoint representation of $I_c(E_{11})$ by acting on it with  $(D_{\underline \alpha} )_C{}^D$ and consider the quantity  $(D^{\underline \alpha} )_C{}^D D_DV^C$ which, using equation (5.8),  transforms as 
$$
(D^{\underline \alpha} )_C{}^D
(D_DV^C)^\prime = (D^{\underline \alpha} )_C{}^D
(D_DV^C)+a_{\underline \beta} (f^{\underline\beta\underline\alpha} {}_{\underline\gamma}-f^{- \underline\beta\underline\alpha}{}_{\underline\gamma})(D^{\underline \gamma}) _C{}^D D_DV^C 
\eqno(5.9)$$
from which it is clear that it does indeed transform in the adjoint representation of $I_C(E_{11})$. In evaluating this equation we  have used the result  
$$
D(h)_D{}^F(D^{\underline \alpha})_F{}^E  D(h^{-1})_E{}^C
= D(hR^{\underline \alpha} h^{-1})_D{}^E = D^{\underline \alpha} 
+a_{\underline \beta}(f^{\underline\beta\underline\alpha} {}_{\underline\gamma}-f^{- \underline\beta\underline\alpha}{}_{\underline\gamma})D^{\underline \gamma}
\eqno(5.10)$$
where $h$ is the infinitesimal transformation given below equation (5.1) and  that by definition $1+D^{\underline \alpha} =D(I+  R^{\underline \alpha})$. 
\par
Examining equation (5.4) we find an additional factor of $(D^{\underline \alpha} )_A{}^B$ in the definition of the covariant derivative. However, when the indices $A$ and $B$ are specified the value of $\underline \alpha $ is determined. As such the effect of multiplying by this factor is just an equivalent way to writing  the quantity 
$(D_{\underline \alpha} )_C{}^D D_DV^C$. Using the fact that the transformation of the adjoint representation simply rearranges the sum over $\underline \alpha$ we find that the covariant derivative defined in equation (5.4) transforms as 
$$
({\cal D}_A V^B)^\prime = D(h^{-1})_A{}^E {\cal D}_E V^F D(h)_F{}^B
\eqno(5.11)$$
When examining this formula it is important to remember that the factors of $D(h)$ are projected by the $(D^{\underline \alpha})_A{}^B $ to lie in the adjoint representation. 
\par
The covariant derivative discussed  in this section has the same form as the one  which appears in the gauge transformations acting on the gauge parameters. This motivates our use of the projector onto the adjoint representation when constructing the covariant derivative, however, as is apparent from equation (5.8) even if we do not project the covariant derivative of equation (5.6) is indeed covariant.  The steps below,  in  conjunction with appendix B,  can be used to evaluate either covariant derivative. 
\par
We will now evaluate in more explicit form the covariant derivative of $V^A$ taking particular components. To do this we will use the equations of appendix B and in particular we repeat  equation (B.16) which is given by 
$$
{\cal D}_a\,V_N 
= D_a\,V_N - 10\,d_{NMP}\,D^M\,V_a{}^P - 12\left(D^\alpha\right)_N{}^M\,D^b{}_M\,V_{ab,\,\alpha} - {2\over 3}\,D^b{}_N\,V_{ab} - {1\over 3}\,D^b{}_N\,V_{ba}+\ldots .
\eqno(5.12)$$
Discarding the covariant derivatives of the level two components of $V^A$ and all higher level components, that is $D^c{}_M\,V_{ab,\,\alpha}=0=D^c{}_N\,V_{ab}=\ldots$ and using the expression for the inverse vielbein of equation (A.32),  and equations (B.3), and (B.10) of appendix B,  we find that 
$$
{\cal D}_a\,V_N 
=  (\det e)^{{1\over 2}}\{ e_a{}^\mu (\partial_\mu V_N +V^c\Omega_{\mu ,}{}_{cN} + 
V_P \Omega_{\mu ,} {}_{\alpha} (D^\alpha -D^{-\alpha})_N{}^P +
10 d_{PMN} V^M_e \Omega _{\mu , }{}^{eP}\dots )
$$
$$
+A_{aP} d^P_{\dot P} (\partial^{\dot P}  V_N +V^c\Omega^{\dot P}{}_{ ,}{}_{cN} + 
V_M \Omega^{\dot P}{}_{ ,} {}_{\alpha} (D^\alpha -D^{-\alpha})_N{}^M +\dots )
$$
$$
+(2 A_{ab}{}^M -{1\over 2} d^{M ST} A_{aS}A_{bT})e_\mu{}^b d_M^{\dot Q} 
(\partial^\mu {}_{\dot Q} V_N +V^c\Omega^{\mu}{}_{\dot Q}{}_{ ,}{}_{cN}+\dots )
$$
$$
-10 d_{NMP} d_{\dot R}^M (\partial^{\dot R}V^P_a +2 V^c\Omega^{\dot R} {}_{, ca}{}^P -V_L\Omega ^{\dot R} {}_{, aQ}d^{QLP} 
$$
$$
-V_a^L \Omega^{\dot R} _{, \alpha} (D^\alpha -D^{-\alpha})_L{}^P +2 V_d^P \Omega^{\dot R}{}_{, a}{}^d+\dots )+\ldots \}
\eqno(5.13)$$
The appearance of the $(\det e)^{{1\over 2}}$ factors will be compensated by the change from a tangent to a world index on $V^A$ which  requires the vielbein and so leads to the factor of $(\det e)^{-{1\over 2}}$. 
 \par
In this last formula we now further truncate and set all components of $V^A$ to zero except for $V_N$ to find that 
$$
{\cal D}_a\,V_N 
=  (\det e)^{{1\over 2}}\{ e_a{}^\mu (\partial_\mu V_N +
V_P \Omega_{\mu ,} {}_{\alpha} (D^\alpha -D^{-\alpha})_N{}^P 
$$
$$
+A_{aP} d^P_{\dot P} (\partial^{\dot P}  V_N + 
V_M \Omega^{\dot P}{}_{ ,} {}_{\alpha} (D^\alpha -D^{-\alpha})_N{}^M  )
+10 d_{NMP} d_{\dot R}^M V_L\Omega ^{\dot R} {}_{, aQ}d^{QLP} 
 +\ldots \}
\eqno(5.14)$$

\par
We can repeat the above process for the derivative of the component $V^b$. The expansion of equation (5.4) when  we take the values $A=a$ and $B=b$ is given in appendix B in particular in equation (B.14). We then carry out truncations in the same spirit as above to find the expression 
$$
{\cal D}_a\,V^b = (\det e)^{{1\over 2}}\Big[e_a{}^\mu\,\left(\partial_\mu\,V^b - 2\,\Omega_{\mu,\,[c}{}^{b]}\,V^c - \Omega_\mu{}^{bM}\,V_M + ...\right)
$$
$$
+\,A_{aN}\,d_{\dot N}{}^N\,\left(\partial^{\dot N}\,V^b - 2\,\Omega^{\dot N}{}_{[c}{}^{b]}\,V^c - \Omega^{{\dot N}\,bM}\,V_M + ...\right)
$$
$$
+\,\left(2\,A_{ac}{}^N - {1\over 2}\,d^{NPR}\,A_{aP}\,A_{cR}\right)\,e_\mu{}^c\,d_N{}^{\dot N}\,\left(\partial^\mu{}_{\dot N}\,V^b - 2\,\Omega^\mu{}_{\dot N,\,[c}{}^{b]}\,V^c - \Omega^\mu{}_{\dot N}{}^{bM}\,V_M + ...\right)
$$
$$
-\,e_\mu{}^b\,d_N{}^{\dot N}\,\left(\partial^\mu{}_{\dot N}\,V_a{}^N + 2\,\Omega^\mu{}_{{\dot N},\,ca}{}^N\,V^c - d^{NMP}\,\Omega^\mu{}_{{\dot N},\,aM}\,V_P + ...\right)\Big]
$$
$$
-\,{1\over 2}\,\left(\det{e}\right)^{1\over 2}\,\delta_a^b\,\Big[e_d{}^\mu\,\left(\partial_\mu\,V^d - 2\,\Omega_{\mu,\,[c}{}^{d]}\,V^c - \Omega_\mu{}^{dM}\,V_M + ...\right)
$$
$$
+\,A_{dN}\,d_{\dot N}{}^N\,\left(\partial^{\dot N}\,V^d - 2\,\Omega^{\dot N}{}_{[c}{}^{d]}\,V^c - \Omega^{{\dot N}\,dM}\,V_M + ...\right)
$$
$$
+\,\left(2\,A_{dc}{}^N - {1\over 2}\,d^{NPR}\,A_{dP}\,A_{cR}\right)\,e_\mu{}^c\,d_N{}^{\dot N}\,\left(\partial^\mu{}_{\dot N}\,V^d - 2\,\Omega^\mu{}_{\dot N,\,[c}{}^{d]}\,V^c - \Omega^\mu{}_{\dot N}{}^{dM}\,V_M + ...\right)
$$
$$
+\,d_{\dot N}{}^N\,\left(\partial^{\dot N}\,V_N + \left(D^\alpha - D^{-\,\alpha}\right)_N{}^M\,\Omega^{\dot N}{}_\alpha + 10\,d_{NMP}\,\Omega^{{\dot N},\,dM}\,V_d{}^P +...\right)
$$
$$
-\,d^{NRS}\,d_R{}^{\dot N}\,A_{cS}\,\left(\partial^c{}_{\dot N}\,V_N + \left(D^\alpha - D^{-\,\alpha}\right)_N{}^M\,\Omega^c{}_{\dot N,\,\alpha} + 10\,d_{NMP}\,\Omega^c{}_{\dot N}{}^{dM}\,V_d{}^P +...\right)
$$
$$
+\,e_\mu{}^d\,d_N{}^{\dot N}\,\left(\partial^\mu{}_{\dot N}\,V_d{}^N + 2\,\Omega^\mu{}_{{\dot N},\,cd}{}^N\,V^c - d^{NMP}\,\Omega^\mu{}_{{\dot N},\,dM}\,V_P + ...\right) + ...\Big]
\eqno(5.16)$$
\par
Proceeding in the same way we find that the covariant derivative of $V_a^N$ is given by 

$$
{\cal D}_a\,V_b{}^N = -2 (\det e)^{{1\over 2}}\,e_{[a}{}^\mu\,\Big(\partial_\mu\,V_{b]}{}^N - d^{NMP}\,\Omega_{\mu,\,b]M}\,V_P
$$
$$
-\,\left(D^\alpha - D^{-\,\alpha}\right)_M{}^N\,\Omega_{\mu,\,\alpha}\,V_{b]}{}^M + 2\,\Omega_{\mu,\,b]}{}^d\,V_d{}^N + 2V^e\Omega _{\mu , eb}{}^N ...\Big) + ...\,.
\eqno(5.17)$$

\medskip
{\bf 6 The world volume viewpoint}
\medskip

The deliberations of the paper so far have largely been based on tangent space quantities.   In this section we will transfer our results to  an approach based on world volume quantities. We begin by reformulating the gauge transformations of equation (3.1) by replacing  the parameter of gauge transformations $\Lambda^A$  by the world volume object $\Lambda^\Pi$, the two quantities being related  
by $\Lambda^A E_A{}^\Pi= \Lambda^\Pi$. We begin by  processing  the gauge transformation of equation (3.1) which can  be written as 
$$
 \delta E_\Pi {}^A= (D^{\underline \alpha})_B{}^A E_\Pi {}^B 
(D^{\underline \alpha})_C{}^D E_D {}^\Sigma  E_\Delta {}^C E_F{}^\Delta D_\Sigma \Lambda ^F 
\eqno(6.1)$$
As explained above equation (2.14), the vielbein can be written as  $E_\Pi {}^A = D(g_E)_\Pi{}^A$. Substituting this into the above equation,  using  arguments similar  to those  given below equation (5.10)  and the fact that $\underline \alpha$ is a dummy index we find that 
$$
 \delta E_\Pi {}^A= (D^{\underline \alpha})_\Pi{}^\Sigma  E_\Sigma {}^A 
(D^{\underline \alpha})_\Sigma {}^\Theta E_C {}^\Sigma  D_\Theta \Lambda ^C
\eqno(6.2)$$
Changing to the world volume parameter this equation becomes 
$$
\delta E_\Pi {}^A= (D^{\underline \alpha})_\Pi{}^\Sigma  E_\Sigma {}^A 
(D^{\underline \alpha})_\Sigma {}^\Theta (\partial_\Theta \Lambda ^\Sigma 
+E_F{}^\Sigma (\partial_\Theta E_\Gamma {}^F -E_\Gamma {}^D \Omega_ {\Theta , }{}_{D}{}^F )\Lambda ^\Gamma )
\eqno(6.3)$$
Using the fact that the torsion tensor is given by 
$$
T_{\Theta \Delta }{}^ C= \partial_\Theta E_\Delta{}^C - \partial_\Delta E_\Theta {}^C- E_\Delta {}^D \Omega_{\Theta , D}{}^C +E_\Theta {}^D \Omega_{\Delta , D}{}^C 
\eqno(6.4)$$
we may rewrite equation (6.3) as  
$$
\delta E_\Pi {}^A= (D^{\underline \alpha})_\Pi{}^\Sigma  E_\Sigma {}^A 
(D^{\underline \alpha})_\Sigma {}^\Theta D_\Theta \Lambda ^\Sigma  + \Lambda^\Gamma \partial_\Gamma E_\Pi{}^A
\eqno(6.5)$$
where 
$$
D_\Theta \Lambda ^\Sigma = \partial_\Theta \Lambda ^\Sigma 
+ T_{\Theta \Gamma }{}^ \Sigma \Lambda^\Gamma - 
E_C {}^\Sigma \Omega_ {\Gamma , }{}_{\Theta}{}^C \Lambda ^\Gamma
\eqno(6.6)$$
\par
In evaluating  equation (6.3) we have used the identity 
$$
(D^{\underline \alpha})_\Pi{}^\Delta  E_\Delta {}^A 
(D^{\underline \alpha})_\Sigma {}^\Theta E_C{}^\Sigma \Lambda ^\Gamma \partial_\Gamma E_\Theta {}^C = 
(D^{\underline \alpha})_F{}^A E_\Pi {}^F (D_{\underline \alpha})_C{}^D \Lambda ^\Gamma G_{\Gamma , D }{}^C
$$
$$
= \Lambda ^ \Gamma \partial_\Gamma E_\Pi {}^ A
\eqno(6.7)$$
In the last step we have adopted the normalisation 
$$
(D^{\underline \alpha})_C{}^D (D_{\underline \beta})_D{}^C =\delta ^{\underline \alpha}_{\underline \beta} 
\eqno(6.8)$$
\par
We note that the part of the  connection in equation (6.3) that is in $I_c(E_{11})$ results in a local $I_c(E_{11})$ transformation of $E_\Pi {}^A$ of the form 
$$
-(D^{\underline \alpha})_F{}^A E_\Pi{}^F (D_{\underline \alpha})_C{}^D 
\Lambda^\Gamma \Omega _{\Gamma , D}{}^C
= - (D^{\underline \alpha}-D^{-\underline \alpha})_F{}^A E_\Pi{}^F
\Lambda^\Gamma \Omega _{\Gamma ,\underline \alpha} 
\eqno(6.9)$$
and so it can be dropped form the variation of $\delta E_\Pi {}^A$ if desired. 
\par
We now demand that the derivative that appears in the transformation of the vielbein of equation (6.5) is just an ordinary derivative, that is, it comes with no connection. The condition for this to be true we require that  
$$
(D^{\underline \alpha}+D^{-\underline \alpha})_\Sigma {}^\Theta(T_{\Theta \Gamma }{}^ \Sigma - 
E_C {}^\Sigma \Omega_ {\Gamma , }{}_{\Theta}{}^C )=0
\eqno(6.10)$$
That we choose the particular  sum over  the values of $\underline \alpha$ in the above expression is related to the fact, noted above,  that the same sum with a minus sign is an $I_c(E_{11})$ transformation.  

As a result the transformation of the vielbein is given by 
$$
E _A{}^\Sigma \delta E_\Pi {}^A= (D^{\underline \alpha})_\Pi{}^\Sigma  
(D^{\underline \alpha})_\Sigma {}^\Theta \partial_\Theta \Lambda ^\Sigma  + \Lambda^\Gamma \partial_\Gamma E_\Pi{}^A E_A{}^\Sigma 
\eqno(6.11)$$
This is the same  pattern that appears in general relativity; the general coordinate transformation of the vielbein $e_\mu{}^a$ contains space-time derivatives of a parameter  that appear with a connection when the parameter has a tangent index,  but with no connection when the parameter has a world index.  As a result the definition of the covariant derivative depends on whether it is acting on a world  parameter or a tangent parameter. The same is true for the generalised geometry we are constructing in this paper. 
\par
In section five we introduced the $l_1$ representation $V^A$ which carried the tangent space index $A$.  We  now consider the corresponding  world volume object  $V^\Pi = V^A E_A{}^\Pi$. Pursuing the analogy with general relativity further we take the vector $V^A$ to transform under the local transformations as 
$$
\delta V^A= \Lambda ^\Gamma \partial _\Gamma V^A  
\eqno(6.12)$$
Using the transformation of the vielbein of equation (6.5), we then find   that under a gauge transformation $V^\Pi$ 
transforms as 
$$
\delta V^{\Pi} = -\,V^\Lambda\,\big(D^{\underline \alpha}\big)_\Lambda{}^\Pi\,\big(D_{\underline \alpha}\big)_\Sigma{}^\Xi\,\partial_\Xi\Lambda^\Sigma + \Lambda^\Sigma\,\partial_\Sigma V^\Pi + \mu\,V^\Pi\,\partial_\Sigma \Lambda^\Sigma.
\eqno(6.13)$$
We have added by hand the last term to take into account of the fact that $V^\Pi$ could be a density. We note that had we not insisted that the transformation  of the vielbein contain an ordinary derivative, rather than a covariant derivative, then the transformation of $V^\Pi$ would not transform into itself alone. 
\par
If we take $\Pi= \dot N$ then  equation (6.13) becomes 
$$
\delta V_{\dot N} = -\,6\,V^{\dot M}\,\big(D^{\alpha}\big)_{\dot N}{}^{\dot M}\,\big(D_{\alpha}\big)_{\dot P}{}^{\dot Q}\,\partial^{\dot P}\Lambda_{\dot Q} + \Lambda^\Sigma\,\partial_\Sigma V_{\dot N} + \mu\,V_{\dot N}\,\partial^{\dot P} \Lambda_{\dot P}.
\eqno(6.14)$$
provided we brutally truncate it to keep only the components $V_{\dot N}$ and $\Lambda^{\dot N}$ of the vectors $V^\Pi$ and $\Lambda ^\Pi$ respectively. This agrees with the transformation of this object given in the papers [35,36] on exceptional field theory.  
\par
We now evaluate the gauge transformations of equation (6.11) for the level zero and one fields. We begin with the transformation of the vielbein which occurs at the lowest level of the generalised vielbein, given in equation (A.28),  in the combination ${\tilde e}_\mu{}^a = \left(\det{e}\right)^{-\,{1\over 2}}\,e_\mu{}^a $. 
Taking  $\Pi = \mu$ and $\Lambda = \nu$ in equation (6.11) we find that the right-hand side becomes 
$$
\delta {\tilde e}_\mu{}^a\,{\tilde e}_a{}^\nu = 
$$
$$
= (D^{\underline \alpha})_\mu{}^\nu  
(D^{\underline \alpha})_\Sigma {}^\Theta \partial_\Theta \Lambda ^\Sigma 
+ \Lambda^\Pi\,\partial_\Pi \tilde e_\mu{}^a\,\tilde e_a{}^\nu
$$
$$
= \,\big(D^\lambda{}_\sigma\big)_\mu{}^\nu\,\left(\delta_\lambda^\eta\,\delta_\rho^\sigma - {1\over 3}\,\delta_\lambda^\sigma\,\delta_\rho^\eta\right)\,\left[\big(D^\rho{}_\eta\big)_\tau{}^\kappa\,\partial_\kappa \Lambda^\tau + \big(D^\rho{}_\eta\big)^{\dot N}{}_{\dot M}\,\partial^{\dot M} \Lambda_{\dot N} + \big(D^\rho{}_\eta\big)_\tau{}^{{\dot N},\,\kappa}{}_{\dot M}\,\partial^\tau{}_{\dot N} \Lambda_\kappa{}^{\dot M}\right]
$$
$$
+\Lambda^\Pi\,\partial_\Pi {\tilde e}_\mu{}^a\,{\tilde e}_a{}^\nu
$$$$
=\,\big(D^\nu{}_\mu\big)_\tau{}^\kappa\,\partial_\kappa \Lambda^\tau + \big(D^\nu{}_\mu\big)^{\dot N}{}_{\dot M}\,\partial^{\dot M} \Lambda_{\dot N} + \big(D^\nu{}_\mu\big)_\tau{}^{{\dot N},\,\kappa}{}_{\dot M}\,\partial^\tau{}_{\dot N} \Lambda_\kappa{}^{\dot M} + \Lambda^\Pi\,\partial_\Pi {\tilde e}_\mu{}^a\,{\tilde e}_a{}^\nu
$$
$$
=\partial_\mu \Lambda^\nu - \partial^\nu{}_{\dot N}\,\Lambda_\mu{}^{\dot N} - {1\over 2}\,\delta_\mu^\nu\,\left(\partial_\sigma \Lambda^\sigma + \partial^{\dot N} \Lambda_{\dot N} + \partial^\sigma{}_{\dot N}\,\Lambda_\sigma{}^{\dot N}\right) + \Lambda^\Pi\,\partial_\Pi {\tilde e}_\mu{}^a\,{\tilde e}_a{}^\nu,
\eqno(6.15)$$
where we have only included the components of the parameter $\Lambda^\Pi$ at levels zero, one and two and discarded the rest. On the other hand the left-hand side of equation (6.11) is given by 
$$
\delta {\tilde e}_\mu{}^a\,{\tilde e}_a{}^\nu = \delta e_\mu{}^a\,e_a{}^\nu - {1\over 2}\,\delta_\mu^\nu\,\big(e_a{}^\lambda\,\delta e_\lambda{}^a\big).
\eqno(6.16)$$
and so  
$$
{\tilde e}_a{}^\lambda\,\delta {\tilde e}_\lambda{}^a = -\,{3\over 2}\,e_a{}^\lambda\,\delta e_\lambda{}^a.
\eqno(6.17)$$
Using this result, and equation (6.15),  we find that 
$$
\delta e_\mu{}^a\,e_a{}^\nu = \delta {\tilde e}_\mu{}^a\,{\tilde e}_a{}^\nu - {1\over 3}\,\delta_\mu^\nu\,\big({\tilde e}_a{}^\lambda\,\delta {\tilde e}_\lambda{}^a\big)
$$
$$
=\,\partial_\mu \Lambda^\nu - \partial^\nu{}_{\dot N}\,\Lambda_\mu{}^{\dot N} + {1\over 3}\,\delta_\mu^\nu\,\partial^{\dot N} \Lambda_{\dot N} + {2\over 3}\,\delta_\mu^\nu\,\partial^\sigma{}_{\dot N}\,\Lambda_\sigma{}^{\dot N} + \Lambda^\Pi\,\partial_\Pi e_\mu{}^a\,e_a{}^\nu.
\eqno(6.18)$$
\par
We now consider the gauge transformations of the level zero scalar fields which are contained in the scalar vielbein $d_{\dot N}{}^N$, however, the scalar vielbein, given in equation (A.28),  contains the combination ${\tilde d}_{\dot N}{}^M = \left(\det{e}\right)^{-\,{1\over 2}}\,d_{\dot N}{}^M$. 
Taking  $\Pi = {\dot N}$ and $\Sigma = {\dot M}$ in equation (6.11) give the equation 
$$
{\tilde d^P{}}_{\dot M}{}\delta {\tilde d}^{\dot N}{}_P  =  
(D^{\underline \alpha})^{\dot N}{}_{\dot M} 
(D^{\underline \alpha})_\Sigma {}^\Theta \partial_\Theta \Lambda ^\Sigma 
+{\tilde d^P{}}_{\dot M}{}\Lambda^\Pi\,\partial_\Pi {\tilde d}^{\dot N}{}_P 
$$
$$
=  6\,\big(D^{\alpha}\big)_{\dot M}{}^{\dot N}\,\big(D_{\alpha}\big)_{\dot Q}{}^{\dot P}\,\left(\partial^{\dot Q}\Lambda_{\dot P} - \partial^\mu{}_{\dot P}\Lambda_\mu{}^{\dot Q}\right)+{\tilde d^P{}}_{\dot M}{}\Lambda^\Pi\,\partial_\Pi {\tilde d}^{\dot N}{}_P 
$$
$$
+\,\big(D^\lambda{}_\sigma\big)_{\dot M}{}^{\dot N}\,\left(\delta_\lambda^\eta\,\delta_\rho^\sigma - {1\over 3}\,\delta_\lambda^\sigma\,\delta_\rho^\eta\right)\,\left[\big(D^\rho{}_\eta\big)_\tau{}^\kappa\,\partial_\kappa \Lambda^\tau + \big(D^\rho{}_\eta\big)^{\dot P}{}_{\dot Q}\,\partial^{\dot Q} \Lambda_{\dot P} + \big(D^\rho{}_\eta\big)_\tau{}^{{\dot P},\,\kappa}{}_{\dot Q}\,\partial^\tau{}_{\dot P} \Lambda_\kappa{}^{\dot Q}\right]
$$
$$
= 6\,\big(D^{\alpha}\big)_{\dot M}{}^{\dot N}\,\big(D_{\alpha}\big)_{\dot Q}{}^{\dot P}\,\left(\partial^{\dot Q}\Lambda_{\dot P} - \partial^\mu{}_{\dot P}\Lambda_\mu{}^{\dot Q}\right) 
-\,{1\over 2}\,\delta_{\dot M}^{\dot N}\,\left(\partial_\mu \Lambda^\mu + {5\over 3}\,\partial^{\dot P} \Lambda_{\dot P} + {7\over 3}\,\partial^\mu{}_{\dot P} \Lambda_\mu{}^{\dot P}\right)
$$
$$
+{\tilde d^P{}}_{\dot M}{}\Lambda^\Pi\,\partial_\Pi {\tilde d}^{\dot N}{}_P \eqno(6.19)$$
While the left-hand side is given by 
$$
{\tilde d^P{}}_{\dot M}{}\delta {\tilde d}^{\dot N}{}_P
 = { d^P{}}_{\dot M}{}\delta { d}^{\dot N}{}_P
 - {1\over 2}\,\delta_{\dot M}^{\dot N}\,\big(e_a{}^\lambda\,\delta e_\lambda{}^a\big).
\eqno(6.20)$$
As a result, truncating as before,  we find that 
$$
\delta d_N{}^{\dot N}\,d_{\dot M}{}^N = 6\,\big(D^{\alpha}\big)_{\dot M}{}^{\dot N}\,\big(D_{\alpha}\big)_{\dot Q}{}^{\dot P}\,\left(\partial^{\dot Q}\Lambda_{\dot P} - \partial^\mu{}_{\dot P}\Lambda_\mu{}^{\dot Q}\right) + { d^P{}}_{\dot M}{}\Lambda^\Pi\,\partial_\Pi { d}^{\dot N}{}_P 
.
\eqno(6.21)$$
\par
In order to find the transformation of level one field $A_{aN}$ we take $\Pi = \mu$ and $\Sigma = {\dot N}$. The left-hand side of equation (6.11) is then given by 
$$
E_{A{\dot N}}\,\delta E_\mu{}^A = E_{a{\dot N}}\,\delta E_\mu{}^a + E^P{}_{\dot N}\,\delta E_{\mu P} = \left(\det{e}\right)^{{1\over 2}}\,A_{a{\dot N}}\,\delta\left[\left(\det{e}\right)^{-\,{1\over 2}}\,e_\mu{}^a\right]
 - d^P{}_{\dot N} \delta A_{\mu P}
$$
$$
=\,-\,\delta A_{\mu{\dot N}} - \left(\delta d_P{}^{\dot M}\,d_{\dot N}{}^P\right)A_{\mu{\dot M}} + \left(e_a{}^\nu\,\delta e_\mu{}^a\right)A_{\nu{\dot N}},
\eqno(6.22)$$
where $\left(\delta d_P{}^{\dot M}\,d_{\dot N}{}^P\right)$ and $\left(e_a{}^\nu\,\delta e_\mu{}^a\right)$ were found in equations (6.21)  and (6.18). While the right-hand side of equation (6.11) is given by 
$$
E_{A{\dot N}}\,\delta E_\mu{}^A = \left(D^{\underline \alpha}\right)_{\mu{\dot N}}\,\left(D_{\underline \alpha}\right)_\Pi{}^\Lambda\,\partial_\Lambda \Lambda^\Pi + \Lambda^\Pi\,\left(E_{A{\dot N}}\,\partial_\Pi E_\mu{}^A\right)
$$
$$
= \left(D_{\mu{\dot N}}\right)^{{\dot M}\nu}\,\partial_\nu \Lambda_{\dot M} + \left(D_{\mu{\dot N}}\right)^\nu{}_{{\dot M},}{}^{\dot P}\,\partial_{\dot P} \Lambda_\nu{}^{\dot M} + \Lambda^\Pi\,\left(E_{A{\dot N}}\,\partial_\Pi E_\mu{}^A\right)+\ldots 
$$
$$
=\,\partial_\mu \Lambda_{\dot N} - 10\,d_{{\dot N}{\dot M}{\dot P}}\,\partial^{\dot M} \Lambda_\mu{}^{\dot P} - \Lambda^\Pi\,\partial_\Pi A_{\mu{\dot N}} - \Lambda^\Pi\,\left(d_{\dot N}{}^N\,\partial_\Pi d_N{}^{\dot M}\right)A_{\mu{\dot M}} + \Lambda^\Pi\,\left(e_a{}^\nu\,\partial_\Pi e_\mu{}^a\right)A_{\nu{\dot N}}+\ldots .
\eqno(6.23)$$
Combining these last two equations  we find that 
$$
\delta A_{\mu{\dot N}} = \Lambda^\Pi\,\partial_\Pi A_{\mu{\dot N}} - \partial_\mu \Lambda_{\dot N} + 10\,d_{{\dot N}{\dot M}{\dot P}}\,\partial^{\dot M}\,\tilde \Lambda_\mu{}^{\dot P}  + \partial_\mu \Lambda^\nu A_{\nu{\dot N}} 
$$
$$
-A_{\mu{\dot M}}\partial^{\dot M} \Lambda_{\dot N}   -10\,d_{{\dot N}{\dot P}{\dot R}}\,d^{{\dot M}{\dot Q}{\dot R}}\Lambda_{\dot Q}\,\partial^{\dot P} A_{\mu{\dot M}} +\ldots ,
\eqno(6.24)$$
In deriving this result we have used the following identity
$$
\,10\,d_{{\dot N}{\dot P}{\dot R}}\,d^{{\dot M}{\dot Q}{\dot R}}\,A_{\mu{\dot M}}\,\partial^{\dot P} \Lambda_{\dot Q} = \,10\,d_{{\dot N}{\dot P}{\dot R}}\,d^{{\dot M}{\dot Q}{\dot R}}\,\partial^{\dot P}\left(\Lambda_{\dot Q}\,A_{\mu{\dot M}}\right) - 10\,d_{{\dot N}{\dot P}{\dot R}}\,d^{{\dot M}{\dot Q}{\dot R}}\,\Lambda_{\dot Q}\,\partial^{\dot P} A_{\mu{\dot M}},
\eqno(6.25)$$
the definition  
$$
\tilde \Lambda_\mu{}^{\dot P} = \Lambda_\mu{}^{\dot P} + d^{{\dot P}{\dot Q}{\dot R}}\,\Lambda_{\dot Q}\,A_{\mu{\dot R}}
\eqno(6.26)$$
and we have discarded level two derivatives 
of the level two parameter, that is, terms of the form $\partial^\nu{}_{\dot M} \bullet$ where $\bullet$ is any parameter. Keeping only the parameter $\Lambda _{\dot N}$, except for one term with the parameter $\Lambda_\mu{}^{\dot P} $, we find that the variation of the level one field is given by 
$$
\delta A_{\mu{\dot N}} = - \partial_\mu \Lambda_{\dot N}   + 
\Lambda^{\dot N}\,\partial_{\dot N} A_{\mu{\dot N}} 
-A_{\mu{\dot M}}\partial^{\dot M} \Lambda_{\dot N}   - 10\,d_{{\dot N}{\dot P}{\dot R}}\,d^{{\dot M}{\dot Q}{\dot R}}\Lambda_{\dot Q}\,\partial^{\dot P} A_{\mu{\dot M}} 
$$
$$+ 10\,d_{{\dot N}{\dot M}{\dot P}}\,\partial^{\dot M}\,\tilde \Lambda_\mu{}^{\dot P} +\ldots 
\eqno(6.27)$$
We note that in carrying out this last step we have, in equation (6.26),  redefined the level two parameter before we have discarded it. The similarly truncated results for the vielbeins in space-time and the scalar sector of equations (6.18) and (6.21) are given respectively by 
$$
\delta e_\mu{}^a\,e_a{}^\nu 
=\,\partial_\mu \Lambda^\nu + {1\over 3}\,\delta_\mu^\nu\,\partial^{\dot N} \Lambda_{\dot N}  + \Lambda^{\dot P}\,\partial_{\dot P} e_\mu{}^a\,e_a{}^\nu.
\eqno(6.28)$$
and 
$$
\delta d_N{}^{\dot N}\,d_{\dot M}{}^N = 6\,\big(D^{\alpha}\big)_{\dot M}{}^{\dot N}\,\big(D_{\alpha}\big)_{\dot Q}{}^{\dot P}\partial^{\dot Q}\Lambda_{\dot P}  + { d^P{}}_{\dot M}{}\Lambda^{\dot P}\,\partial_{\dot P} {d}^{\dot N}{}_P .
\eqno(6.29)$$

Equations (6.27-29) are indeed the the transformations of the gauge fields found in the papers [35,36] on the five dimensional exceptional field theory. However, to get these formulae we have carried out a brutal from the $E_{11}$ perspective. In particular,  we have set all components of $\Lambda ^\Pi$ to zero except $\Lambda _{\dot N}$, something which is only compatible with the symmetry $GL(5)\otimes E_6$ part of the $E_{11}$ symmetry. 

\medskip
{\bf {7. Conclusion}}
\medskip
In this paper we have started from the $E_{11}$ approach in five dimensions  and evaluated, at the low levels,  the corresponding gauge transforms and generalised field strengths for the  fields which are  defined on a generalised space-time whose coordinates are the usual coordinates  of space-time as well as Lorentz scalar coordinates that belong to the ${\bar {27}}$-dimensional representation of $E_{6}$.  We also introduced matter fields transforming in the $l_1$ representation  into the non-linear realisation and gave their transformation rules. After a radical truncation the resulting formulae  agree  with those proposed in exceptional field theory. As we explained in the introduction the main ideas that underlie    exceptional field theory are those proposed much earlier in the context of the $E_{11}\otimes_s l_1$ non-linear realisation and, as one can choose to work at low levels in the later,  it should not come as too great a surprise to find that one does indeed recover exceptional field theory after a truncation. While we have not computed all possible quantities the reader can be left in no doubt that exceptional field theory is  a truncation of the $E_{11}\otimes_s l_1$ non-linear realisation. 
\par
The truncation required to find exceptional field theory only preserves  the $E_{6}$ symmetry and breaks all the higher  level $E_{11}$ symmetries. However, if one would like to extend exceptional field theory to include 
larger symmetries one only has to include the higher level fields and coordinates in the $E_{11}\otimes_s l_1$ non-linear realisation, that is, 
adopt the $E_{11}\otimes_s l_1$ non-linear realisation as ones starting point. Seen in its proper context exceptional field theory provides evidence for the $E_{11}$ approach. 
\par
As we have mentioned in the $E_{11}$ approach the dynamics is determined using the symmetries of the $E_{11}\otimes_s l_1$ non-linear realisation. This was carried out in four dimensions for the gauge fields and scalars [22], although the calculation of  the gravity-dual gravity equation was not fully worked out. Analogous result were also found in eleven dimensions [21]. It would be straightforward to find the equivalent  results in five dimensions and it would be very surprising if these, when radically truncated,  do not agree with those of exceptional field theory. We note that we do not apparently need the gauge transformations studied in this paper and certainly  not the matter fields. However, it could be that a knowledge of the gauge transformations could shorten the calculation. 
\par
One fact that emerged in references [22,21] was that the $E_{11}\otimes_s l_1$ invariant equations of motion were invariant under the usually understood gauge transformations of form fields and we see in reference [24] and section four that, at the linearised level, they are also invariant under the low level gauge transformations associated with the generalised space-time. Thus it might appear that the symmetries of the $E_{11}\otimes_s l_1$ imply the presence of the diffeomorphism and gauge symmetries. This possibility requires further study. 
\par
To find the dynamics of the  $E_{11}\otimes_s l_1$ non-linear realisation at 
higher levels requires us to find  equations of motion for the fields which are beyond the usual supergravity fields. However, these fields have in general  mixed space-time indices, that is they do not just possess one block of totally antisymmetrised indices.  As such they possess gauge transformations that are much more complicated, but are correctly given by the gauge transformations proposed in reference [24] where the gauge parameters are in one to one correspondence with the $l_1$ representation.
As a result the  gauge invariant field equations for these higher and higher level fields will involve more and more space-time derivatives, indeed one more space-time derivative for each additional index block. It follows that the equations of motion cannot be given by  a Ricci tensor equal zero condition as this has only two derivatives. The gauge transformations contain the rigid $E_{11}$ transformations [24], but these symmetries will rotate the low level supergravity fields into the higher level mixed index fields and as a result one might  not expect the Riemann tensor to be gauge invariant.  As such the geometry that the non-linear realisation corresponds to must be quite different to that which encodes Einstein's theory and more similar to that found in higher spin theories. 
\par
The dynamics of $E_{11}\otimes_s l_1$ non-linear realisation of the four and eleven  dimensional theories given in references [22,21] is formulated in terms of duality equations which are first order in derivatives. However, these equations only hold modulo gauge transformations  and to eliminate these one must act on the equations with more and more derivatives as the number of gauge transformations grows. This is consistent with the remarks made in the previous paragraph. 
\par
Exceptional field is defined so as to possess a section condition which restricts the dependence of the fields on the generalised space-time. Indeed this condition is required from the very outset in order to perform  calculations in this theory. However, in the $E_{11}$ approach one does not, at least up to now,  need such a condition to find the equations of motion. As such it is possible that the need for the section condition in exceptional field theory is a result of the radical truncation required to find exceptional field theory in the $E_{11}$ approach. As was pointed out in reference [47] the section condition of Siegel theory is a BPS condition and it is not usual to impose such a condition from the beginning. How the restrictions on the field dependence on the generalised space-time might arise was addressed from the viewpoint of the first quantise string theory in reference [48]. One can think of the generalised space-time introduced in the $E_{11}\otimes_s l_1$ non-linear realisation as an effective spacetime just as one thinks about a field theory as an effective field theory. From this perspective how to recover the spacetime we are used to is a physical question whose resolution might require us to understand the physics of spacetime in  a more fundamental way. 
\par
To find the different theories in the $E_{11}$ approach one decomposes the algebra into different subalgebras. As such the theories in the different dimensions are the same but the quantities are rearranged [10]. Hence it is obvious from this perspective that the theory in five, or any other, dimensions will lift, or descend, to the theory in the new  dimension. 
\par
As we mentioned in the introduction the $E_{11}$ conjecture is that the effective theory of string and branes is given by the dynamics encoded in the $E_{11}\otimes_s l_1$ non-linear realisation.  If  one has could show that the dynamics,  when truncated to the low level  fields which depend on the usual spacetime only, are just those of maximal supergravity theory then one would, in effect,   have to believe in the conjecture. This follows  as some of the fields present at higher levels in the $E_{11}$ theory are known to present in the underlying theory. In reference [22] this result was shown for the  vector and scalar fields and the graviton equation was also correct at linearised order [49]. The technical problem that arose for  the graviton equation is now understood [51], using the ideas put forward in reference [50],  and it is to be hoped that this calculation can now be completed. 

\medskip
{\bf {Acknowledgements}}
\medskip
We wish to thank the SFTC for support from Consolidated grant number ST/J002798/1 and Alexander Tumanov wishes to thanks King's College  for the support provided by his  Graduate School International Research Studentship. 
\medskip

\magnification1200

\centerline {\bf {Appendix A: The five dimensional theory} }
\medskip
In this appendix we review the $E_{11}\otimes_s l_1$ non-linear realisation in five dimensions. The five dimensional theory is obtained by deleting node 5 from the $E_{11}$ Dynkin diagram to find the algebra $Gl(5)\otimes E_6$. The $E_{11} \otimes_s l_1$ algebra is then decomposed into representations of this  subalgebra of $E_{11}$ [12,44].
$$
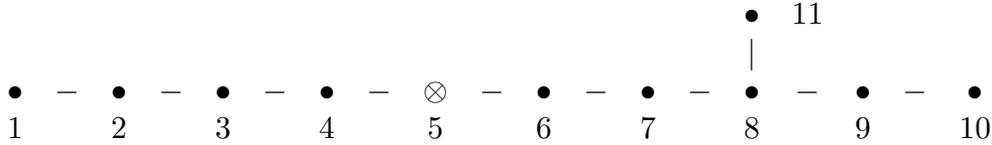
\matrix{
& & & & & & & & & & & & & & \bullet & 11 & & & \cr 
& & & & & & & & & & & & & & | & & & & \cr
\bullet & - & \bullet & - & \bullet & - & \bullet & - & \otimes & - & \bullet & - & \bullet & - & \bullet & - & \bullet & - & \bullet \cr
1 & & 2 & & 3 & & 4 & & 5 & & 6 & & 7 & & 8 & & 9 & & 10 \cr
}
$$
The generators of $E_{11}$ are denoted by $R^{\underline \alpha}$. The level of an $E_{11}$ generators is just the number of up minus down $GL(5)$ indices. The positive, including zero, level generators of the $E_{11}$ up to level 3 are
$$
K^a{}_b,\ R^{\alpha}, \  R^{aM}, \  R^{a_1a_2}{}_{N}, \ R^{a_1a_2a_3,\alpha}, \ R^{a_1a_2,\,b} \ldots 
\eqno(A.1)
$$
where $R^{[a_1a_2,\,b]} = 0$. The negative level generators are given by 
$$
R_{aM}, \  R_{a_1a_2}{}^{N}, \ R_{a_1a_2a_3}{}^\alpha, \   R_{a_1a_2,\,b}, \ldots 
\eqno(A.2)
$$ 
\par
The $l_1$ representation decomposes into representations of $Gl(5)\otimes E_6$ as follows 
$$
l_A=\{ P_a, \quad Z^{N}, \ Z^{a}{}_{N}, \ Z^{a_1a_2,\,\alpha}, \ Z^{ab} \ldots  \}
\eqno(A.3)
$$
The fourth generator $Z^{ab}$ has no symmetries on its indices. The level is the number of up minus down GL(5) indices plus one. For all these objects the lower case Latin indexes correspond to 5-dimensional fundamental representation of $GL(5)$ ($a,\,b,\,c,\,... = 1,\,...,\,5$). Lower case Greek indexes correspond to 78-dimensional adjoint representation of $E_6$ ($\alpha,\,\beta,\,\gamma,\,... = 1,\,...,\,78$). We note that the generators of $E_6$ are denoted by $R^{\alpha}$ which has no underline on the $\alpha$. to distinguish it from the generators of $E_{11}$.
Upper and lower case Latin indexes correspond to ${\overline 27}$-dimensional and ${27}$-dimensional representations of $E_6$ ($N,\,M,\,P,\,... = 1,\,...,\,27$) respectively. An  arbitrary group element  of $E_{11}\otimes_s l_1$ can be written as  $g=g_l g_E$ where 
$$
g_l = \exp{\left(x^a\,P_a + x_{N}\,Z^{N} + x_{a}{}^{N}\,Z^{a}{}_{N} + x_{a_1a_2,\,\alpha}\,Z^{a_1a_2,\,\alpha} + x_{ab}\,Z^{ab}+\ldots\right )}, 
$$
$$
g_E = \exp{\left(h_{a}{}^{b}\,K^{a}{}_{b}\right)}\,\exp{\left(\varphi_\alpha\,R^\alpha\right)}\,\exp{\left(A_{a_1a_2a_3,\,\alpha}\,R^{a_1a_2a_3,\,\alpha}\right)} 
$$
$$
\times \exp{\left(A_{a_1a_2,\,b}\,R^{a_1a_2,\,b}\right)}\,\exp{\left(A_{a_1a_2}{}^{N}\,R^{a_1a_2}{}_{N}\right)}\,\exp{\left(A_{aN}\,R^{aN}\right)\ldots }. \eqno(A.4)
$$
In writing this group element we have used the local symmetry of the non-linear realisation $g\to gh$ to gauge away all terms that involve negative level generators. The group element $g_l$ is labelled by the quantities  
$$
x^a, \ x_{N}, \ x_{a}{}^{N}, \ x_{a_1a_2,\alpha}, \ x_{ab}, \ \ldots
\eqno(A.5)$$ 
which will be identified with the generalised coordinates of the generalised space-time, while the fields
$$
h_a{}^b, \ \varphi_{\alpha}, \ A_{aM}, \ A_{a_1a_2}{}^{N}, \ A_{a_1a_2a_3,\,\alpha}, \ A_{a_1a_2,\,b}, \ \ldots
\eqno(A.6)$$
which depend on the generalised space-time.
\par
We now give the $E_{11}$ algebra when written in terms of the above generators. The commutators of the $E_{11}$ generators with the GL(4) generators $K^a{}_b$ are
$$
\left[K^a{}_b,\,K^c{}_d \right] = \delta^{c}_{b}\,K^a{}_d - \delta^{a}_{d}\,K^c{}_b, \quad \left[K^a{}_b,\,R^\alpha \right] = 0,
$$
$$
\left[K^a{}_b,\,R^{cN} \right] = \delta^{c}_{b}\,R^{aN},\quad \left[K^a{}_b,\,R_{cN} \right] = -\,\delta^{a}_{c}\,R_{bN},
$$
$$
\left[K^a{}_b,\,R^{a_1a_2}{}_{N} \right] = 2\,\delta^{[a_1}_{\,b}\,R^{|a|a_2]}{}_{N}, \quad \left[K^a{}_b,\,R_{a_1a_2}{}^{N} \right] = -\,2\,\delta^{\,a}_{[a_1}\,R_{|b|a_2]}{}^{N},
$$
$$
\left[K^a{}_b,\,R^{a_1a_2a_3,\,\alpha} \right] = 3\,\delta^{[a_1}_{\,b}\,R^{|a|a_2a_3],\,\alpha}, \quad \left[K^a{}_b,\,R_{a_1a_2a_3}{}^\alpha \right] = -\,3\,\delta^{\,a}_{[a_1}\,R_{|b|a_2a_3]}{}^\alpha, 
$$
$$
\left[K^a{}_b,\,R^{a_1a_2,\,c} \right] = 2\,\delta^{[a_1}_{\,b}\,R^{|a|a_2],\,c} + \delta^c_b\,R^{a_1a_2,\,a}, \quad \left[K^a{}_b,\,R_{a_1a_2,\,c} \right] = -\,2\,\delta^{\,a}_{[a_1}\,R_{|b|a_2],\,c} - \delta^a_c\,R_{a_1a_2,\,b}.
$$
$$
\left[K^a{}_b,\,R^{a_1...a_4}{}_{N_1N_2} \right] = 4\delta^{[a_1}_{\,b}\,R^{|a|a_2a_3a_4]}{}_{N_1N_2}, \quad \left[K^a{}_b R_{a_1...a_4}{}^{N_1N_2} \right] = -\,4\,\delta^{\,a}_{[a_1}\,R_{|b|a_2a_3a_4]}{}^{N_1N_2},
$$
$$
\left[K^a{}_b,\,R^{a_1a_2a_3,\,cN} \right] = 3\,\delta^{[a_1}_{\,b}\,R^{|a|a_2a_3],\,cN} + \delta^{c}_{b}\,R^{a_1a_2a_3,\,aN},
$$
$$
\left[K^a{}_b,\,R_{a_1a_2a_3,\,cN}\right] = -\,3\,\delta^{\,a}_{[a_1}\,R_{|b|a_2a_3],\,cN} - \delta_c^a\,R_{a_1a_2a_3,\,bN}. \eqno(A.7)
$$
\par
The commutators of the $E_{11}$ generators with $E_6$ generators $R^\alpha$ are determined by the representation of $E_6$ that this generator belongs to. They are given by 
$$
\left[R^\alpha,\,R^\beta\right] = f^{\alpha\beta}{}_{\gamma}\,R^\gamma, \quad \left[R^\alpha,\,R^{aM}\right] = (D^\alpha)_N{}^M R^{aN}, \quad \left[R^\alpha,\,R_{aM}\right] = -\,\left(D^\alpha\right)_M{}^N R_{aN},
$$
$$
\left[R^\alpha,\,R^{a_1a_2}{}_M\right] = -\,(D^\alpha)_M{}^N R^{a_1a_2}{}_N, \quad \left[R^\alpha,\,R_{a_1a_2}{}^N\right] = \left(D^\alpha\right)_M{}^N R_{a_1a_2}{}^M,
$$
$$
\left[R^\alpha,\,R^{a_1a_2a_3,\,\beta}\right] = f^{\alpha\beta}{}_{\gamma}\,R^{a_1a_2a_3,\,\gamma}, \quad \left[R^\alpha,\,R_{a_1a_2a_3}{}^\beta\right] = f^{\alpha\beta}{}_{\gamma}\,R_{a_1a_2a_3}{}^{\gamma},
$$
$$
\left[R^\alpha,\,R^{a_1a_2,\,b}\right] = 0, \quad \left[R^\alpha,\,R_{a_1a_2,\,b}\right] = 0, 
$$
$$
\left[R^\alpha,\,R^{abcd}{}_{MN}\right]= -\,(D^\alpha)_M{}^P R^{abcd}{}_{PN} - (D^\alpha)_N{}^P R^{abcd}{}_{MP},
$$
$$
\left[R^\alpha,\,R_{abcd}{}^{MN}\right] = \left(D^\alpha\right)_P{}^M R^{abcd}{}^{PN} + \left(D^\alpha\right)_P{}^N R_{abcd}{}^{MP}, \eqno(A.8)
$$
$$
\left[R^\alpha,\,R^{a_1a_2a_3,\,bM}\right] = (D^\alpha)_N{}^M R^{a_1a_2a_3,\,bN}, \quad \left[R^\alpha,\,R_{a_1a_2a_3,\,bM}\right] = -\,\left(D^\alpha\right)_M{}^N R_{a_1a_2a_3,\,bN}.
$$
where $f^{\alpha\beta}{}_{\gamma}$ are the structure constants of $E_6$, normalised by
$$
 f_{\alpha\beta\gamma}\,f^{\alpha\beta\delta} = -\,4\,\delta^\delta_\gamma
\eqno(A.9)$$
We lower and raise indices on the $E_6$ generators with the Cartan-Killing metric $g_{\alpha\beta}$ of $E_6$. The matrices   $(D^\alpha)_N{}^M$ are the generators of $E_6$ in ${\bf 27}$ representation and so obey the relation 
$$
  [D^\alpha , D^\beta ]_M{}^N = f^{\alpha\beta}{}_\gamma (D^\gamma )_M{}^N 
\eqno(A.10)$$
They are normalised so that $$
  (D^\alpha)_M{}^N (D^\beta)_N{}^M = g^{\alpha \beta}
\eqno(A.11)$$ 
\par
The commutation relations of the positive level $E_{11}$ generators are given by
$$
\left[R^{a M},\,R^{b N}\right] = d^{MNP}\,R^{ab}{}_P, \quad \left[R^{a N},\,R^{bc}{}_M\right] = \left(D_\alpha \right)_M{}^N R^{abc,\,\alpha} + \delta^N_M\,R^{bc,a}, 
$$
$$
\left[R^{ab}{}_M,\,R^{cd}{}_N\right] = R^{abcd}{}_{MN} - 20\,d_{MNP}\,R^{ab[c,\,d]P}, \quad \left[R^{aN},\,R^{bc,\,d}\right] = R^{abc,\,dN} - {1\over 3}\,R^{bcd,\,aN},
$$
$$
\left[R^{aN},\,R^{bcd,\,\alpha}\right] = 3\,d^{NMP}\,\left(D^\alpha\right)_P{}^R R^{abcd}{}_{MR} + 6\,\left(D^\alpha\right)_M{}^N R^{bcd,\,aM}. \eqno(A.12)
$$
where $d^{MNP}$ is the completely symmetric invariant tensor of
$E_6$ formed from the product of three $\overline {\bf 27}$ representations. The invariant tensor $d_{MNP}$, which has its indices down,  is 
completely symmetric product of three ${\bf 27}$ indices and 
satisfies the relation 
$$
  d^{MNP} d_{MNQ} = \delta^P_Q 
\eqno(A.13)$$
We follow the conventions of reference [44]  rather than reference [12], the difference being a rescaling of $d$ by $\sqrt{5}$. 
We will also use  the relation [44] 
$$
  g_{\alpha \beta} D^\alpha_M{}^N D^\beta_P{}^Q = {1 \over 6} \delta^N_P
  \delta_M^Q + {1 \over 18} \delta_M^N \delta_P^Q -{5 \over 3} d^{NQR} d_{MPR}
\eqno(A.14)$$
\par
The commutators of negative-level $E_{11}$ generators are 
$$
[R_{aN},\,R_{bM} ] = d_{NMP}\,R_{ab}{}^P, \quad [R_{aN},\,R_{bc}{}^M ] = \left(D_\alpha\right)_N{}^M\,R_{abc}{}^\alpha + \delta _N^M\,R_{bc,a}.
$$
$$
\left[R_{ab}{}^M,\,R_{cd}{}^N\right] = R_{abcd}{}^{MN} - 20\,d^{MNP}\,R_{ab[c,\,d]P}, \quad \left[R_{aN},\,R_{bc,\,d}\right] = R_{abc,\,dN} - {1\over 3}\,R_{bcd,\,aN},
$$
$$
\left[R_{aN},\,R_{bcd,\,\alpha}\right] = 3\,d_{NMP}\,\left(D_\alpha\right)_R{}^P R_{abcd}{}^{MR} + 6\,\left(D_\alpha\right)_N{}^M R_{bcd,\,aM}. \eqno(A.15)
$$
The commutators between the positive and negative level generators of $E_{11}$ up to level 4 are given by
$$
\left[R^{aN},\,R_{bM}\right] = 6\,\delta_a^b\,(D_\alpha)_M{}^N R^\alpha + \delta_M^N\,K^a{}_b - {1\over 3}\,\delta_M^N\,\delta^a_b\,K^c{}_c,
$$
$$
\left[R_{aN},\,R^{bc}{}_{M}\right] = 20\,d_{NMP}\,\delta_{\,a}^{[b}\,R^{c]P}, \quad \left[R^{aN},\,R_{bc}{}^{M}\right] = 20\,d^{NMP}\,\delta^{\,a}_{[b}\,R_{c]P},
$$
$$
\left[R_{aN},\,R^{a_1a_2a_3,\,\alpha}\right] = 18\left(D^\alpha\right)_N{}^M \delta^{[a_1}_{\,a}R^{a_2a_3]}{}_{M}, \quad \left[R^{aN},\,R_{a_1a_2a_3}{}^\alpha\right] = 18\left(D^\alpha\right)_M{}^N \delta_{[a_1}^{\,a}R_{a_2a_3]}{}^{M},
$$
$$
\left[R_{aN},\,R^{a_1a_2,\,b}\right] = \delta^{b}_{a}\,R^{a_1a_2}{}_{N} - \delta^{[b}_{\,a}\,R^{a_1a_2]}{}_{N}, \quad \left[R^{aN},\,R_{a_1a_2,\,b}\right] = \delta_{b}^a\,R_{a_1a_2}{}^{N} - \delta_{[b}^{\,a}\,R_{a_1a_2]}{}^{N},
$$
$$
\left[R_{aN},\,R^{a_1...a_4}{}_{N_1N_2}\right] = 40\,d_{N[N_1|M|}\,\left(D_\alpha\right)_{N_2]}{}^M \delta_{\,a}^{[a_1}\,R^{a_2a_3a_4],\,\alpha},
$$
$$
\left[R^{aN},\,R_{a_1...a_4}{}^{N_1N_2}\right] = 40\,d^{N[N_1|M|}\,\left(D_\alpha\right)_{M}{}^{N_2]}\,\delta^{\,a}_{[a_1}\,R_{a_2a_3a_4}{}^{\alpha}, \eqno(A.16)
$$
$$
\left[R_{aN},\,R^{a_1a_2a_3,\,bM}\right] = \left(D_\alpha\right)_N{}^M \delta_a^b\,R^{a_1a_2a_3,\,\alpha} - \left(D_\alpha\right)_N{}^M \delta_{\,a}^{[b}\,R^{a_1a_2a_3],\,\alpha} + 3\,\delta_N^M\,\delta_{\,a}^{[a_1}\,R^{a_2a_3],\,b},
$$
$$
\left[R^{aN},\,R_{a_1a_2a_3,\,bM}\right] = \left(D_\alpha\right)_M{}^N \delta^a_b\,R_{a_1a_2a_3}{}^{\alpha} - \left(D_\alpha\right)_M{}^N \delta^{\,a}_{[b}\,R_{a_1a_2a_3]}{}^{\alpha} + 3\,\delta_M^N\,\delta^{\,a}_{[a_1}\,R_{a_2a_3],\,b}.
$$
The Cartan involution acts on the generators of $E_{11}$ as follows
$$
I_c\left(K^a{}_b\right) = -\,K^b{}_a, \quad I_c\left(R^\alpha\right) = -\,R^{-\alpha}, \quad I_c\left(R^{aN}\right)= -\,J^{MN}\,R_{aM},
$$
$$
I_c\left(R^{ab}{}_M\right)= J^{-1}_{MN}\,R_{ab}{}^N ,\quad I_c\left(R^{abc,\,\alpha}\right) = -\,R_{abc,\,-\alpha}, \quad I_c\left(R^{a_1a_2,\,c}\right) = -\,R_{a_1a_2,\,c},
$$
$$
I_c\left(R^{abcd}{}_{MN}\right) = J^{-1}_{MP}\,J^{-1}_{NQ}\,R_{abcd}{}^{PQ}, \quad I_c\left(R^{abc,\,dN}\right) = J^{NM}\,R_{abc,\,dM}. \eqno(A.17)
$$
where the constant $J_{NM}$ relates generators with the {\bf 27} and ${\bf \bar {27}}$ representation under the Cartan involution. 
\par
The scalar product on the $E_{11}$ algebra $(R^{\underline \alpha} , R^{\underline \beta} )\equiv g^{{\underline \alpha}{\underline \beta}}$
defines the Cartan-Killing metric $g^{{\underline \alpha}{\underline \beta}}$. It is straightforward to calculate using its  invariance property,  $([R^{\underline \gamma }, R^{\underline \alpha}] , R^{\underline \beta} )+ (R^{\underline \alpha} , [R^{\underline \gamma }, R^{\underline \beta}] )=0$ together with the above commutation relations. 
\par 
We find that the  $E_{11}$ Killing metric in the decomposition suitable to five dimensions is found to be given  up to level 2 by 
$$
g^{{\underline \alpha}{\underline \beta}} = \left(\matrix{
{1\over 6}\,g^{\alpha\beta} & 0 & 0 & 0 & 0 & 0 \cr
0 & \delta^a_d\,\delta_b^c - {1\over 2}\,\delta^a_b\,\delta_d^c & 0 & 0 & 0 & 0 \cr
0 & 0 & 0 & \delta^a_b\,\delta^N_M & 0 & 0 \cr
0 & 0 & \delta_a^b\,\delta_N^M & 0 & 0 & 0 \cr
0 & 0 & 0 & 0 & 0 & -\,20\,\delta_N^M\,\delta_{b_1b_2}^{a_1a_2} \cr
0 & 0 & 0 & 0 & -\,20\,\delta^N_M\,\delta^{b_1b_2}_{a_1a_2} & 0 \cr
}\right), \eqno(A.18)
$$
where the generators in the scalar product are labelled by 
$$R^{\underline \alpha} = \{ R^\alpha,\,K^a{}_b,\,R^{aN},\,R_{aN},\,R^{a_1a_2}{}_N,\,R_{a_1a_2}{}^N,\,...\}
\eqno(A.19)$$ 
and  
$$R^{\underline \beta} = \{ R^\beta,\,K^c{}_d,\,R^{cM},\,R_{cM},\,R^{b_1b_2}{}_M,\,R_{b_1b_2}{}^M,\,...\}$$. 
The next entries for the metric are more easily given by listing the scalar product 
$$
\left( R^{a_1a_2a_3,\,\alpha},\,R_{b_1b_2b_3}{}^\beta \right) = 360\,g^{\alpha\beta}\,\delta_{b_1b_2b_3}^{a_1a_2a_3}, \quad \left( R^{a_1a_2,\,a},\,R_{b_1b_2,\,b} \right) = 20\,\delta_b^a\,\delta_{b_1b_2}^{a_1a_2} - 20\,\delta_{b\,b_1b_2}^{a\,a_1a_2}.
\eqno(A.20)$$
To avoid confusion of notation we denote $g^{\alpha\beta}$ to be  the Killing metric on $E_6$ and we never write the $\alpha, \beta$ components of the $E_{11}$ Cartan-Killing metric, although of course we do use it. 
\par
We now give the commutators between the generators of $E_{11}$ and those of the $l_1$ representation. The commutation relations between the later and the generators of GL(5) are given by
$$
\left[K^a{}_b,\,P_c\right] = -\,\delta^a_c\,P_b + {1\over 2}\,\delta^a_b\,P_c, \quad \left[K^a{}_b,\,Z^N\right] = {1\over 2}\,\delta^a_b\,Z^N, \quad \left[K^a{}_b,\,Z^c{}_N\right] = \delta^c_b\,Z^a_N + {1\over 2}\,\delta^a_b\,Z^c_N,
$$
$$
\left[K^a{}_b,\,Z^{a_1a_2,\,\alpha}\right] = 2\,\delta^{[a_1}_{\,b}Z^{|a|a_2],\,\alpha} + {1\over 2}\,\delta^a_b\,Z^{a_1a_2,\,\alpha}, \quad \left[K^a{}_b,\,Z^{cd}\right] = \delta^c_b\,Z^{ad} + \delta^d_b\,Z^{ca} + {1\over 2}\,\delta^a_b\,Z^{cd}. \eqno(A.21)
$$
while with the generators of $E_6$ we have
$$
\left[R^\alpha,\,P_a\right] = 0, \quad \left[R^\alpha,\,Z^M\right] = (D^\alpha)_N{}^M Z^N, \quad \left[R^\alpha,\,Z^a{}_N\right] = -\,\left(D^\alpha\right)_N{}^M Z^a{}_M,
$$
$$
\left[R^\alpha,\,Z^{a_1a_2,\,\beta}\right] = f^{\alpha\beta}{}_\gamma\,Z^{a_1a_2,\,\gamma}, \quad \left[R^\alpha,\,Z^{ab}\right] = 0. \eqno(A.22)
$$
\par

The elements of the $l_1$ representation at a given level can be introduced into the algebra by taking the commutators of  suitable $E_{11}$ generators of the same level with $P_a$, namely 
$$
\left[R^{aN},\,P_b\right] = \delta_b^a\,Z^N, \quad \left[R^{a_1a_2}{}_{N},\,P_a\right] = 2\,\delta_{a}^{[a_1}Z^{a_2]}{}_N,
$$
$$
\left[R^{a_1a_2a_3,\,\alpha},\,P_a\right] = 3\,\delta_{\,a}^{[a_1}Z^{a_2a_3],\,\alpha}, \quad \left[R^{a_1a_2,\,b},\,P_a\right] = -\,2\,\delta_a^b\,Z^{[a_1a_2]} - 2\,\delta_{\,a}^{[a_1}\,Z^{|b|a_2]}, \eqno(A.23)
$$
The commutators of the remaining positive level generators of $E_{11}$ with the $l_1$ generators is  determined by the Jacobi identities and they are found to be given by 
$$
\left[R^{aM},\,Z^N\right] = -\,d^{MNP}\,Z^a{}_P, \quad \left[R^{aN},\,Z^b{}_M\right] = -\,\left(D_\alpha\right)_M{}^N Z^{ab,\,\alpha} - \delta^N_M\,Z^{ab},
$$
$$
\left[R^{a_1a_2}{}_{N},\,Z^M\right] = -\,\left(D_\alpha\right)_N{}^M Z^{a_1a_2,\,\alpha} + 2\,\delta^N_M\,Z^{[a_1a_2]}. \eqno(A.24)
$$
Commutators between the level $-1$ generators of $E_{11}$ and those of the $l_1$ representation are also determined by the Jacobi identities to be given by 
$$
\left[R_{aN},\,P_b\right] = 0, \quad \left[R_{aN},\,Z^M\right] = \delta_N^M\,P_a, \quad \left[R_{aN},\,Z_M^b\right] = -\,10\,d_{NMP}\,\delta_a^b\,Z^P,
$$
$$
\left[R_{aN},\,Z^{a_1a_2,\,\alpha}\right] = -\,12\,\left(D^\alpha\right)_N{}^M \delta_{\,a}^{[a_1}Z^{a_2]}{}_M, \quad \left[R_{aN},\,Z^{bc}\right] = -\,{2\over 3}\,\delta_a^b\,Z^c{}_N - {1\over 3}\,\delta_a^c\,Z^b{}_N. \eqno(A.25)
$$
Commutators with level $-2$ generators are
$$
\left[R_{a_1a_2}{}^N,\,P_b\right] = 0, \quad \left[R_{a_1a_2}{}^N,\,Z^M\right] = 0, \quad \left[R_{a_1a_2}{}^N,\,Z^b{}_M\right] = 20\,\delta_M^N\,\delta^{\,b}_{[a_1}\,P_{a_2]},
$$
$$
\left[R_{a_1a_2}{}^N,\,Z^{b_1b_2,\,\alpha}\right] = 120\,\delta_{a_1a_2}^{b_1b_2}\,\left(D^\alpha\right)_M{}^N\,Z^M, \quad \left[R_{a_1a_2}{}^N,\,Z^{b_1b_2}\right] = -\,{20\over 3}\,\delta_{a_1a_2}^{b_1b_2}\,Z^N. \eqno(A.26)
$$
Commutators with level $-3$ generator $R_{a_1a_2a_3}{}^\alpha$ are
$$
\left[R_{a_1a_2a_3}{}^\alpha,\,P_b\right] = 0, \quad \left[R_{a_1a_2a_3}{}^\alpha,\,Z^M\right] = 0, \quad \left[R_{a_1a_2a_3}{}^\alpha,\,Z^b{}_M\right] = 0,
$$
$$
\left[R_{a_1a_2a_3}{}^\alpha,\,Z^{b_1b_2,\,\beta}\right] = 360\,g^{\alpha\beta}\,\delta_{[a_1a_2}^{\,b_1b_2}\,P_{a_3]}, \quad \left[R_{a_1a_2a_3}{}^\alpha,\,Z^{bc}\right] = 0. \eqno(A.27)
$$
\par
The generalised vielbein is defined in equation (2.11) and it was found in references [12,46] to be given by 
$$
E_\Pi{}^A = \left( \det{{} e} \right)^{-\,{1\over 2}} 
$$ 
$$
\def\quad{\hskip1ex\relax}
\left(\matrix{
{} e_\mu^{\,\,\,a} & {} e_\mu{}^{b}\,\alpha_{b|M} & {} e_\mu{}^{b}\,\alpha_{b|a}{}^M & {} e_\mu{}^{b}\,\alpha_{b|a_1a_2,\,\alpha} & {} e_\mu{}^{b}\,\alpha_{b|cd} \cr
0 & \left( {} d^{-1} \right)_M{}^{\dot N} & \left( {} d^{-1} \right)_P{}^{\dot N} \beta^{P}{}_a{}^M & \left({}  d^{-1} \right)_P{}^{\dot N} \beta^P{}_{a_1a_2,\alpha} & \left({}  d^{-1} \right)_P{}^{\dot N} \beta^P{}_{cd} \cr
0 & 0 & {} d_{\dot N}{}^M\left({} e^{-1}\right)_a{}^\mu & {} d_{\dot N}{}^P\left({} e^{-1}\right)_b{}^\mu \gamma^b{}_{P|a_1a_2,\,\alpha} & {} d_{\dot N}{}^P\left({} e^{-1}\right)_b{}^\mu \gamma^{b}{}_{P|cd} \cr
0 & 0 & 0 & \left( {} e^{-1} \right)^{\ \mu_1\mu_2}_{a_1a_2}\,\left( f^{-1} \right)_\beta{}^\alpha & 0 \cr
0 & 0 & 0 & 0 & \left( {} e^{-1} \right)_c{}^{\mu}\,\left( {} e^{-1} \right)_d{}^{\nu} \cr
}\right), \eqno(A.28)
$$
where in the first line 
$$
\alpha_{a|N} = -\,A_{aN}, \quad \alpha_{a|b}{}^N = -\,2\,A_{ab}{}^N - {1\over 2}\,d^{NMP}\,A_{aM}\,A_{bP},
$$
$$
\alpha_{a|a_1a_2,\,\alpha} = -\,3\,A_{aa_1a_2,\,\alpha} + 2\,A_{a[a_1}{}^N A_{a_2]M}\,\left(D_\alpha\right)_N{}^M + {1\over 6}\,A_{aN}\,A_{[a_1M}\,A_{a_2]P}\,d^{NMS}\,\left(D_\alpha\right)_S{}^P,
$$
$$
\alpha_{a|cd} = -\,4\,A_{d(a,\,c)} - 2\,A_{ad}{}^N\,A_{cN} - {1\over 6}\,A_{aN}\,A_{bM}\,A_{cP}\,d^{NMP}, \eqno(A.29)
$$
in the second line 
$$
\beta^N{}_a{}^M  = A_{aP}\,d^{NMP}, \quad \beta^N{}_{a_1a_2,\alpha} = A_{a_1a_2}{}^M \left(D_\alpha\right)_M{}^N - {1\over 2}\,A_{[a_1M}\,A_{a_2]R}\,d^{NMP}\left(D_\alpha\right)_P{}^R,
$$
$$
\beta^N{}_{ab} = -\,2\,A_{ab}{}^N + {1 \over 2}\,A_{aM}\,A_{bP}\,d^{NMP}, \eqno(A.30)
$$
and in the third line 
$$
\gamma^a{}_{N|a_1a_2,\,\alpha} = -\,\delta_{[a_1}^{\,a}\,A_{a_2]M}\,\left(D_\alpha\right)_N{}^M, \quad \gamma^a{}_{N|cd} = \delta_d^a\,A_{cN}. 
\eqno(A.31)$$
The indices $\dot N, \dot M,\ldots $ are curved $E_6$ indices while the indices $N, M,\ldots $ are tangent $E_6$ indices. The field ${} e_\mu{}^a = (e^{{} h})_\mu {}^a$, but it will turn out that the actual vielbein is given by $e_\mu{}^a= (\det {} e) ^{-{1\over 2}} {} e_\mu{}^a$ and we will in the main text of the paper use $d_M{}^{\dot N}= (\det {} e) ^{-{1\over 2}}{} d_M{}^{\dot N}$
\par
The inverse generalised veibein is given by 
$$
E_A{}^\Pi = \left( \det{{} e} \right)^{{1\over 2}} 
$$
$$
\def\quad{\hskip1ex\relax}
\left(\matrix{
{} e_a{}^\mu & d_{\dot N}{}^N\,\alpha_{a|N} & \left( {} d^{-1} \right)_M{}^{\dot M}\,{} e_\mu{}^{b}\,\alpha_{a|b}{}^M & {} e_{\mu_1\mu_2}^{\,\,b_1b_2}\,f_{\dot\alpha}{}^{\beta}\,\alpha_{a|b_1b_2,\,\beta} &{}  e_\mu{}^e\,{}  e_\nu{}^f\,\alpha_{a|ef} \cr
0 & {} d_{\dot N}{}^N & \left( {} d^{-1} \right)_M{}^{\dot N}\,{} e_\mu{}^b\,\beta^{N}{}_b{}^M & {} e_{\mu_1\mu_2}^{\,\,b_1b_2}\,f_{\dot\alpha}{}^{\beta}\,\beta^N{}_{a_1a_2,\alpha} & {} e_\mu{}^e\,{} e_\nu{}^f\,\beta^N{}_{cd} \cr
0 & 0 & \left({}  d^{-1} \right)_{N}{}^{\dot N}\,{} e_\mu{}^a & {} e_{\mu_1\mu_2}^{\,\,b_1b_2}\,f_{\dot\alpha}{}^{\beta}\,\gamma^a{}_{N|b_1b_2,\,\beta} &{}  e_\mu{}^e\,{} e_\nu{}^f\,\gamma^{a}{}_{N|ef} \cr
0 & 0 & 0 & {} e_{\mu_1\mu_2}^{\,\,a_1a_2}\,f_{\dot\alpha}{}^{\alpha} & 0 \cr
0 & 0 & 0 & 0 & {} e_\mu{}^c\,{} e_\nu{}^d \cr
}\right),
\eqno(A.32)$$
where in the first line 
$$
\alpha_{a|N} = A_{aN}, \quad \alpha_{a|b}{}^N = 2\,A_{ab}{}^N - {1\over 2}\,d^{NMP}\,A_{aM}\,A_{bP},
\eqno(A.33)$$
$$
\alpha_{a|a_1a_2,\,\alpha} = 3\,A_{aa_1a_2,\,\alpha} + 2\,A_{a[a_1}{}^N A_{a_2]M}\,\left(D_\alpha\right)_N{}^M - {1\over 6}\,A_{aN}\,A_{[a_1M}\,A_{a_2]P}\,d^{NMS}\,\left(D_\alpha\right)_S{}^P,
\eqno(A.34)$$
$$
\alpha_{a|cd} = 4\,A_{d(a,\,c)} - 2\,A_{ad}{}^N\,A_{cN} + {1\over 6}\,A_{aN}\,A_{bM}\,A_{cP}\,d^{NMP},
\eqno(A.35)$$
in the second line 
$$
\beta^N{}_a{}^M  = -\,A_{aP}\,d^{NMP}, \quad \beta^N{}_{a_1a_2,\alpha} = -\,A_{a_1a_2}{}^M \left(D_\alpha\right)_M{}^N + {1\over 2}\,A_{[a_1M}\,A_{a_2]R}\,d^{NMP}\left(D_\alpha\right)_P{}^R,
\eqno(A.36)$$
$$
\beta^N{}_{ab} = 2\,A_{ab}{}^N + {1 \over 2}\,A_{aM}\,A_{bP}\,d^{NMP},
\eqno(A.37)$$
and in the third line 
$$
\gamma^a{}_{N|a_1a_2,\,\alpha} = \delta_{[a_1}^{\,a}\,A_{a_2]M}\,\left(D_\alpha\right)_N{}^M, \quad \gamma^a{}_{N|cd} = -\,\delta_d^a\,A_{cN}.
\eqno(A.38)$$


\magnification1200

\centerline {\bf {Appendix B:  Calculation of the covariant derivatives} }
\medskip

The gauge transformations of section three and the matter fields of section four  both involve covariant derivatives given in equations (3.1) and (5.4) respectively. Although these formulae are very elegant to evaluate them for specific cases can be rather and in this appendix we show in detail how do this is done. 
The covariant derivative in these sections acts on an object, denoted here by $\bullet$,  that  transforms under $I_c(E_{11})$. It consists of a projector acting on a more familiar  type of derivative which is of the generic form 
$$
D_A \bullet = E^{-1}{}_A{}^\Pi \{ \partial _\Pi - \Omega_{\Pi ,\underline \alpha }(D(R^{\underline \alpha})-D(R^{-\underline \alpha}) \}\bullet 
\eqno(B.1)$$
where $\Omega _{\Pi ,}{}_{\underline \alpha}$ is the  $I_C(E_{11})$  valued connection and $D(R^{\underline \alpha})$ is the representation of $R^{\underline \alpha}$ appropriate to the object on which it acts. We recall that by definition 
$ D(1+R^{\underline \alpha})= 1+ D^{\underline \alpha}$ and so we adopt the shorthand notation  $ D(R^{\underline \alpha})= D^{\underline \alpha}$ as this makes it clearer what generators are involved. 
\par
As explained in section five  for the object $V^A$, which transforms in the $l_1$ representation restricted to $I_c(E_{11})$,  the covariant derivative before it is project into the adjoint representation is given by 
$$
D_A V^B= E^{-1}{}_A{}^\Pi (\partial _\Pi V^B- V^E\Omega _{\Pi ,}{}_E{}^B )
\eqno(B.2)$$
where $\Omega _{\Pi ,}{}_E{}^B = \Omega_{\Pi ,\underline \alpha }(D^{\underline \alpha}-D^{-\underline \alpha} )_E{}^B= \Omega_{\Pi ,\underline \alpha }(D(R^{\underline \alpha})-D(R^{-\underline \alpha}) )_E{}^B$. 
\par
Let us begin by evaluating  the inverse generalised vielbein. Using equation (A.32) we find that 
$$
E^{-1}{}_{a\Vert}{}^\Pi\,\partial_\Pi = \left(\det{e}\right)^{1\over 2}\,\Big[e_a{}^\mu\,\partial_\mu + A_{aN}\,d_{\dot N}{}^N\,\partial^{\dot N} + \left(2\,A_{ab}{}^N - {1\over 2}\,d^{NMP}\,A_{aR}\,A_{bS}\right)\,e_\mu{}^b\,d_M{}^{\dot M}\,\partial^\mu{}_{\dot M}+\ldots \Big],
\eqno(B.3)$$
$$
E^{-1} {}^{N\Vert \Pi}\,\partial_\Pi = \left(\det{e}\right)^{1\over 2}\,\Big[d_{\dot N}{}^N\,\partial^{\dot N} - d^{NMP}\,A_{aP}\,e_\mu{}^a\,d_M{}^{\dot M}\,\partial^\mu{}_{\dot M}+\ldots \Big], 
\eqno(B.4)$$
and 
$$
 E^{-1}{}^a{}_N{}^{\Pi}\,\partial_\Pi = \left(\det{e}\right)^{1\over 2}\,d_N{}^{\dot N}\,e_\mu{}^a\,\partial^\mu{}_{\dot N}+\ldots ,
\eqno(B.5)$$
The evaluation of the inverse generalised vielbein in front of the term in the covariant derivative which contains the connection takes an analogous form,  but with the partial derivative exchanged for  the connection. 
\par
We now consider  the evaluation of the connection which has the form 
$$
\Omega _{\Pi ,}{}_E{}^B = \Omega_{\Pi ,\underline \alpha }(D(R^{\underline \alpha})-D(R^{-\underline \alpha}) )_E{}^B= \Omega_{\Pi ,\underline \alpha }(D^{\underline \alpha}-D^{-\underline \alpha} )_E{}^B
$$
$$
= \Omega_{\Pi , }{}_e{}^f(D(K^e{}_{f})-D(K^f{}_e) )_E{}^B 
+ \Omega_{\Pi , }{}_\alpha (D(R^{\alpha})-D(R^{-\alpha}) )_E{}^B 
$$
$$
+ \Omega_{\Pi , }{}_{cP}(D(R^{cP})-D(R_{cP}) )_E{}^B 
+ \Omega_{\Pi , }{}_{e_1e_2}{}^{P}(D(R^{e_1e_2}{}_{P})-D(R_{e_1e_2}{}^{P}) )_E{}^B +\ldots 
\eqno(B.6)$$
Contracting with $V^E$ and evaluating using  the components of the $V^A$ multiplet  of equation (5.3) we find that 
$$
   V^E\Omega _{\Pi ,}{}_E{}^B= V^a \Omega _{\Pi ,}{}_{a\Vert }{}^B
+V_M \Omega _{\Pi ,}{}^M{}_{\Vert }{}^B+ V_c^M \Omega _{\Pi ,}{}_M^c{}_{\Vert }{}^B+\ldots 
$$
$$
= V^a \Omega _{\Pi , \underline \alpha }(D(R^{\underline \alpha})-D(R^{-\underline \alpha}) )_{a\Vert }{}^B
+V_M \Omega _{\Pi , \underline \alpha }(D(R^{\underline \alpha})-D(R^{-\underline \alpha}) )^M{}_{\Vert }{}^B
$$
$$
+ V_c^M \Omega _{\Pi , \underline \alpha }(D(R^{\underline \alpha})-D(R^{-\underline \alpha}) )_M^c{}_{\Vert }{}^B
+
V_{c_1c_2\alpha} \Omega _{\Pi , \underline \beta }(D(R^{\underline \beta})-D(R^{-\underline \beta}) )^{c_1c_2\alpha}{}_{\Vert }{}^B
$$
$$
+V_{c_1c_2} \Omega _{\Pi , \underline \alpha }(D(R^{\underline \alpha})-D(R^{-\underline \alpha}) )^{c_1c_2}{}_{\Vert }{}^B
+\ldots 
\eqno(B.7)$$
When  $B$ takes the value $b$ to find that  equation (B.7) is given by 
$$
 V^E \Omega_{\Pi ,} {}_E{}^b= V^c\Omega_{\Pi , }{}_e{}^f(D(K^e{}_f)-D(K^f{}_e) )_{c\Vert}{}^b 
+ V_M\Omega_{\Pi , }{}_{cP} (D(R^{cP})-D(R_{cP}) )^{M\Vert b} 
$$
$$
+ V_{f \Vert }{}^M \Omega_{\Pi , }{}_{e_1e_2}{}^P(D(R^{e_1e_2}{}_P)-D(R_{e_1e_2}{}^P) )^{f}{}_{M\Vert }{}^{b}
+ \ldots 
\eqno(B.8)$$
and as a result we find, using the algebra of appendix A,  that 
$$
-V^E\Omega _{\Pi ,} {}_E{}^b= -2V^{c} \Omega_ {\Pi ,}{}_{c}{}^b
- V_P \Omega _{\Pi ,}{}^{b P}
-20 V^M_e \Omega_ {\Pi ,}{}^{eb}{}_{ M}+\ldots 
\eqno(B.9)$$
\par
When $B$ takes the value $N$ we find that equation (B.7) takes the form 
$$
-V^E \Omega_{\Pi ,} {}_{E\Vert }{}_N= V^{c} \Omega_ {\Pi ,}{}_{c N}
+ V_M \Omega _{\Pi ,\alpha }(D^\alpha - D^{-\alpha})_N{}^M
+10 d_{PMN } V^M_e \Omega_ {\Pi ,}{}^{e  P}
$$
$$
-120 V_{c_1c_2\beta} \Omega_ {\Pi ,}{} ^{c_1c_2}{}_P{} (D^\beta)_N{}^P
+{20\over 3} V_{c_1c_2} \Omega_ {\Pi ,}{} ^{c_1c_2}{}_N 
+\ldots 
 \eqno(B.10)$$
While when $B$ takes the value ${}_a{}^N$ we find that equation (B.7) takes the form 
$$
-V^E \Omega_{\Pi ,} {}_{E\Vert }{}_{a}{}^N= 2V^{c} \Omega_ {\Pi ,}{}_{ca}{}^N
- V_M \Omega _{\Pi ,aP }d^{PMN} - V^M_a \Omega_ {\Pi ,\alpha }(D^\alpha - D^{-\alpha})_M{}^N 
$$
$$
+2 V_d{}^N  \Omega_ {\Pi ,}{} _{a}{}^d +\dots 
 \eqno(B.11)$$

\par
The covariant derivative of $V^A$ of equation (5.4) and the gauge transformation of equation (3.1) has a projector onto the adjoint representation of $I_C(E_{11})$. We will now show how to evaluate this projector. We called equation (5.4) for the covariant derivative of the vector $V^A$;   
$$
{\cal D}_a\,V^b = \left(D^{\underline \alpha}\right)_a{}^b\,\left(D_{\underline \alpha}\right)_C{}^D\,D_D\,V^C = \left(D^e{}_f\right)_a{}^b\,(D_e{}^f)_C{}^D\,D_D\,V^C.
\eqno(B.12)$$
The first step is to evaluate the projector which using the Cartan-Killing form of equation (A.18),  we find that 
$$
\left(D^e{}_f\right)_A{}^B\,(g^{-1})_f{}^{e,\,m}{}_n\,\left(D^n{}_m\right)_C{}^D = \left(D^e{}_f\right)_A{}^B\,(D^f{}_e)_C{}^D - {1\over 3}\,\left(D^e{}_e\right)_A{}^B\,(D^f{}_f)_C{}^D.
\eqno(B.13)$$
Substituting this into equation (B.12) we find that it becomes 
$$
{\cal D}_a\,V^b = \left(D^e{}_f\right)_a{}^b\,(D^f{}_e)_C{}^D\,D_D\,V^C - {1\over 3}\,\left(D^e{}_e\right)_a{}^b\,(D^f{}_f)_C{}^D\,D_D\,V^C = \left(D^b{}_a\right)_C{}^D\,D_D\,V^C
$$
$$
= D_a\,V^b - D^b{}_N\,V_a{}^N - {1\over 2}\,\delta^a_b\,\left(D_c\,V^c + D^N V_N+ D^c{}_N\,V_c{}^N+\dots \right).
\eqno(B.14)$$
\par
Finally,  we consider the covariant derivative of the $V_N$ component of $V^A$ which is equal to 
$${\cal D}_a\,V_N = \left(D^{\underline \alpha}\right)_{a\Vert N}\,\left(D_{\underline \alpha}\right)_C{}^D\,D_D\,V^C = \left(D^{cP}\right)_{a\Vert N}(D_{cP})_C{}^D\,D_D\,V^C
\eqno(B.15)$$
since the  only value of ${\underline \alpha}$ for which $\left(D^{\underline \alpha}\right)_{aN}$ is non-zero is ${\underline \alpha} = bM$. Using that $\left(D^{bM}\right)_{aM} = -\,\delta_a^b\,\delta_N^M$ and the Cartan-Killing metric of equation (A.) we find 
$$
{\cal D}_a\,V_N = -\,\left(D_{aN}\right)_C{}^D\,D_D\,V^C
$$
$$
= D_a\,V_N - 10\,d_{NMP}\,D^M\,V_a{}^P - 12\left(D^\alpha\right)_N{}^M\,D^b{}_M\,V_{ab,\,\alpha} - {2\over 3}\,D^b{}_N\,V_{ab} - {1\over 3}\,D^b{}_N\,V_{ba}.
\eqno(B.16)$$
Applying the same procedures to the  covariant derivative of the $V_a{}^N$ 
we find that         
$${\cal D}_a\,V_b{}^N = \left(D^{\underline \alpha}\right)_{a,\,b}{}^N\,\left(D_{\underline \alpha}\right)_C{}^D\,D_D\,V^C = -\,{1\over 20}\,\left(D^{a_1a_2}{}_M\right)_{a,\,b}{}^N\,\left(D_{a_1a_2}{}^M\right)_C{}^D\,D_D\,V^C, 
$$
$$
={1\over 10} (D_{ab}{}^N)_C{}^D D_D V^C
\eqno(B.17)$$
where ${}^{a_1a_2}{}_M$ is the only value of ${\underline \alpha}$ that gives a non-zero contribution and $-\,{1\over 20}$ factor comes from the Cartan-Killing metric. This in turn is equal to 
$$
{\cal D}_a\,V_b{}^N = +2\,D_{[a}\,V_{b]}{}^N - 12\,\left(D^\alpha\right)_M{}^N\,D^M\,V_{ab,\,\alpha} + {2\over 3}\,D^N\,V_{[ab]}.
\eqno(B.18)$$

\medskip
{\bf {References}}
\medskip
\item{[1]} S.\ Ferrara, J.\ Scherk and B.\ Zumino, 
{\it Algebraic Properties of Extended Supersymmetry},
Nucl.\ Phys.\ {\bf B121} (1977) 393;
E.\ Cremmer, J.\ Scherk and S.\ Ferrara, {\it SU(4) Invariant
Supergravity Theory}, Phys.\ Lett.\ {\bf 74B} (1978) 61.
\item{[2]} E. Cremmer and B. Julia,
{\it The $N=8$ supergravity theory. I. The Lagrangian},
Phys.\ Lett.\ {\bf 80B} (1978) 48
\item{[3]} B.\ Julia, {\it Group Disintegrations},
in {\it Superspace \&
Supergravity}, p.\ 331,  eds.\ S.W.\ Hawking  and M.\ Ro\v{c}ek,
Cambridge University Press (1981); 
B. Julia, in {\it Vertex Operators in Mathematics  and Physics},
Publications of the Mathematical Sciences Research Institute no 3,  
Springer Verlag (1984); N. Marcus and J. Schwarz, {\it Three-dimensional
supergravity theories}, Nucl. Phys. {\bf B228} (1983) 301; 
H. Nicolai, {\it The integrability of N=16  
supergravity}, Phys. Lett. {\bf 194B} (1987) 402; {\it On M-Theory}, hep-th/9801090. 
\item{[4]} J, Schwarz and P. West,
{\it ``Symmetries and Transformation of Chiral
$N=2$ $D=10$ Supergravity''},
\item{[5]} P. West, {\it $E_{11}$ and M Theory}, Class. Quant.  
Grav.  {\bf 18}, Phys. Lett. {\bf 126B} (1983) 301.
(2001) 4443, hep-th/ 0104081; 
\item{[6]} P. West, {\it Hidden superconformal symmetries of  
M-theory}, {\bf JHEP 0008} (2000) 007, {\tt arXiv:hep-th/0005270}.
\item{[7]} F. Englert, L. Houart, A. Taormina and  P. West, {\it The Symmetry of M-Theories},  JHEP 0309 (2003) 020,  hep-th/0304206. 
\item{[8]} P. West, {\it $E_{11}$, SL(32) and Central Charges},
Phys. Lett. {\bf B 575} (2003) 333-342,  hep-th/0307098. 
\item{[9]} I. Schnakenburg and  P. West, {\it Kac-Moody   
symmetries of
IIB supergravity}, Phys. Lett. {\bf B517} (2001) 421, hep-th/0107181.
\item{[10]} P. West, {\it The IIA, IIB and eleven dimensional theories 
and their common
$E_{11}$ origin}, Nucl. Phys. B693 (2004) 76-102, hep-th/0402140. 
\item{[11]}  F. ÊRiccioni and P. West, {\it
The $E_{11}$ origin of all maximal supergravities}, ÊJHEP {\bf 0707}
(2007) 063; ÊarXiv:0705.0752.
\item{[12]} ÊF. Riccioni and P. West, {\it E(11)-extended spacetime
and gauged supergravities},
JHEP {\bf 0802} (2008) 039, ÊarXiv:0712.1795.
\item{[13]}  A. Kleinschmidt and P. West, {\it  Representations of G+++
and the role of space-time},  JHEP 0402 (2004) 033,  hep-th/0312247.
\item{[14]} P. Cook and P. West, {\it Charge multiplets and masses
for E(11)}, ÊJHEP {\bf 11} (2008) 091, arXiv:0805.4451.
\item{[15]} P. West,  {\it $E_{11}$ origin of Brane charges and U-duality
multiplets}, JHEP 0408 (2004) 052, hep-th/0406150. 
\item{[16]} P. West, {\it Brane dynamics, central charges and
$E_{11}$}, hep-th/0412336. 
\item {[17]} C. Hillmann, {\it Generalized E(7(7)) coset dynamics and D=11
supergravity}, JHEP {\bf 0903}, 135 (2009), hep-th/0901.1581. 
\item{[18]} C. Hillmann, {\it E(7(7)) and d=11 supergravity }, PhD
thesis,  arXiv:0902.1509.
\item{[19]} D.  Berman, H. Godazgar, M. Perry and  P.  West, 
{\it Duality Invariant Actions and Generalised Geometry}, 
arXiv:1111.0459. 
\item{[20]}
D.~S.~Berman, M. Perry, {\it Generalized Geometry and M theory}, 
ÊÊJHEP {\bf 1106} (2011) 74
ÊÊ[arXiv:1008.1763 [hep-th]]; 
D.~S.Berman, H. Godazgar and M. Perry,
{\it SO(5,5) duality in M-theory and generalized geometry}, 
ÊÊPhys.\ Lett.\ Ê{\bf B700 } (2011) Ê65-67.
ÊÊ[arXiv:1103.5733 [hep-th]].
\item {[21]} P. West, {\it Generalised Geometry, eleven dimensions
and $E_{11}$}, JHEP 1202 (2012) 018, arXiv:1111.1642.  
\item{[22]} P. West, {\it  E11, Generalised space-time and equations of motion in four dimensions}, JHEP 1212 (2012) 068, arXiv:1206.7045. 
\item{[23]}  A. Borisov and V. Ogievetsky,  {\it Theory of dynamical affine and conformal  symmetries as the theory of the gravitational field}, 
Teor. Mat. Fiz. 21 (1974) 32. 
\item{[24]} P. West, {\it Generalised Space-time and Gauge Transformations },  arXiv:1403.6395. 
\item{[25]} M. Duff, {\it Duality Rotations In String Theory},
  Nucl.\ Phys.\  B {\bf 335} (1990) 610; M. Duff and J. Lu,
 Duality rotations in
membrane theory,  Nucl. Phys. {\bf B347} (1990) 394; A. Tseytlin, {\it Duality Symmetric Formulation Of String World
Sheet Dynamics}, Phys.Lett. {\bf B242} (1990) 163, {\it Duality Symmetric
Closed String Theory And Interacting Chiral Scalars}, Nucl.\ Phys.\ B {\bf
350}, 395 (1991). 
\item{[26]} W. Siegel, {\it Two vielbein formalism for string inspired axionic gravity},   Phys.Rev. D47 (1993) 5453,  hep-th/9302036; 
\item{[27]} W. Siegel,{\it Superspace duality in low-energy superstrings}, Phys.Rev. D48 (1993) 2826-2837, hep-th/9305073; 
{\it Manifest duality in low-energy superstrings},  
In *Berkeley 1993, Proceedings, Strings '93* 353,  hep-th/9308133. 
\item{[28]} C. Hull and B. Zwiebach, Double Field Theory,  JHEP {\bf 0909}
(2009) 099, hep-th/0904.4664.
\item{[29]}  C. Hull and B. Zwiebach, The gauge algebra of double field
theory and Courant brackets,
  JHEP {\bf 0909} (2009) 090, hep-th0908.1792.
\item{[30]} O. Hohm, C. Hull and B. Zwiebach, Background independent
action for double field theory, hep-th/1003.5027.
\item{[31]} O. Hohm, C. Hull and B. Zwiebach, Generalised metric
formulation of double field theory,  hep-th/1006.4823. 
\item{[32]} O. Hohm and S.  Kwak, {\it Frame-like Geometry of Double Field Theory},   J.Phys.A44 (2011) 085404, arXiv:1011.4101. 
\item{[33]}  P. West, {\it E11, generalised space-time and IIA string
theory}, 
 Phys.Lett.B696 (2011) 403-409,   arXiv:1009.2624.
\item{[34]}   A. Rocen and P. West,  {\it E11, generalised space-time and
IIA string theory;  the R-R sector},  arXiv:1012.2744.
\item{[35]}  O. Hohm and  H. Samtleben, {\it Exceptional Form of D=11 Supergravity}, Phys. Rev. Lett. 111 (2013)  231601, arXiv:1308.1673.
\item{[36]}  O. Hohm and  H. Samtleben, {\it Exceptional Field Theory I: $E_{6(6)}$ covariant Form of M-Theory and Type IIB},  Phys. Rev. D 89, (2014) 066016 , arXiv:1312.0614. 
\item{[37]}  O. Hohm and  H.  Samtleben, {\it Exceptional Field Theory II: E$_{7(7)}$} ,  arXiv:1312.0614. 
\item{[38]} H. Godazgar, M. Godazgar, O. Hohm, H.  Nicolai, Henning Samtleben, {\it Supersymmetric E$_{7(7)}$ Exceptional Field Theory},  
arXiv:1406.3235. 
\item{[39]} O. Hohm and  H. Samtleben, {\it   Exceptional Field Theory III: E$_{8(8)}$},  Phys. Rev. D 90, (2014) 066002, arXiv:1406.3348
\item{[40]} Seminar by P. West in Lyon during the period of  10-14 October 2011 during which H. Samtleben was present and further private discussions with same. 
\item{[41]} D. Berman, M. Cederwall, A. Kleinschmidt and D. Thompson, {\it The gauge structure of generalised diffeomorphisms }, JHEP {\bf 1301} (2013) 064, hep-th:1208.5884; G.  Aldazabal, M. Grana, D.  MarquŽs, and J. Rosabalar, 
{\it The gauge structure of Exceptional Field Theories and the tensor hierarchy}, Xiv:1312.4549; 
A. Coimbra, C. Strickland-Constable and  D.  Waldram,  {\it $E_{d(d)} \times {R}^+$ Generalised Geometry, Connections and M theory}, hep-th:1112.3989.
\item{[42]} P. West, Introduction to Strings and Branes, Cambridge
University Press, June 2012. 
\item{[43]} F. Englert qnd  L. Houart, {\it G+++ Invariant Formulation of Gravity and M-Theories: Exact BPS Solutions }, JHEP0401 (2004) 002,  arXiv:hep-th/0311255.    
\item{[44]} F. Riccioni, D.  Steele and  P. West, {\it The E(11) origin of all maximal supergravities - the hierarchy of field-strengths }, JHEP 0909 (2009) 095,  arXiv:0906.1177. 
\item{[45]} S. Coleman, J. Wess and  B. Zumino, {\it Structure of phenomenological Lagrangians. 1.}, Phys.Rev. 163 (1967) 1727; 
C. Callan, S. Coleman, J. Wess and  B. Zumino,  {\it Structure of phenomenological Lagrangians. 2.}, Phys.Rev. 177 (1969) 2239. 
\item{[46]} A. Tumanov and  P. West, {\it Generalised vielbeins and non-linear realisations },  arXiv:1405.7894. 
\item{[47]} P. West, {\it Generalised BPS conditions }, arXiv:1208.3397.  
\item{[48]} P. West,  {\it   Generalised space-time and duality},    Phys.Lett.{\bf B693} (2010) 373, arXiv:1006.0893. 
\item{[49]} P. West,  {\it  Dual gravity and E11},  arXiv:1411.0920
\item{[50]} N. Boulanger, Per Sundell and P.  West, {\it Gauge fields and infinite chains of dualities}, arXiv:1502.07909. 
\item{[51]} To be published,  but explained by P. West at the conference "La charme discret de la Symetrie",  March 2015,  Brussels.

\end